%% file: final_manuscript.tex
\documentclass[12pt, draftclsnofoot, onecolumn]{IEEEtran}
\usepackage{cite}
\usepackage{graphicx,amsmath,amssymb}
\usepackage{subfigure}
\usepackage{citesort}
\usepackage{fancyhdr}
\usepackage{mdwmath}
\usepackage{mdwtab}
\usepackage{balance}
\usepackage{xcolor}
\usepackage{bm}

\usepackage{algorithm}
\usepackage{multirow}
\usepackage{flafter}
\usepackage{comment}

\usepackage{amsthm}
\usepackage{algpseudocode}

\usepackage{stmaryrd}

\usepackage[font=small]{caption}
\usepackage{float}
\floatstyle{plaintop}
\restylefloat{table}

\newtheorem{remark}{Remark}
\newtheorem{theorem}{Theorem}
\newtheorem{lemma}{Lemma}
\newtheorem{model}{Model}
\newtheorem{proposition}{Proposition}

\newtheorem{corollary}{\textbf{Corollary}}

\allowdisplaybreaks
\begin{document}


\title{STAR-RISs: A Correlated T\&R Phase-Shift Model and Practical Phase-Shift Configuration Strategies}

%

\author{Jiaqi\ Xu~\IEEEmembership{Graduate Student Member,~IEEE}, Yuanwei\ Liu~\IEEEmembership{Senior Member,~IEEE}, Xidong\ Mu~\IEEEmembership{Graduate Student Member,~IEEE}, Robert Schober~\IEEEmembership{Fellow,~IEEE}, and H. Vincent Poor~\IEEEmembership{Life Fellow,~IEEE}.

\thanks{J. Xu and Y. Liu are with the School of Electronic Engineering and Computer Science, Queen Mary University of London, London E1 4NS, UK. (email:\{jiaqi.xu, yuanwei.liu\}@qmul.ac.uk).}
\thanks{X. Mu is with School of Artificial Intelligence, Beijing University of Posts and Telecommunications, Beijing, China. (email: muxidong@bupt.edu.cn).}
\thanks{R. Schober is with the Institute for Digital Communications, Friedrich-Alexander-University Erlangen-N\"{u}rnberg (FAU), Germany (e-mail: robert.schober@fau.de).}
\thanks{H. V. Poor is with the Department of Electrical Engineering, Princeton University, Princeton, NJ 08544 USA (e-mail:
poor@princeton.edu).}
}
\maketitle
\begin{abstract}
A \emph{correlated} transmission and reflection (T\&R) phase-shift model is proposed for passive lossless simultaneously transmitting and reflecting reconfigurable intelligent surfaces (STAR-RISs).
A STAR-RIS-aided two-user downlink communication system is investigated for both orthogonal multiple access (OMA) and non-orthogonal multiple access (NOMA).
To evaluate the impact of the correlated T\&R phase-shift model on the communication performance, three phase-shift configuration strategies are developed, namely the primary-secondary phase-shift configuration (PS-PSC), the diversity preserving phase-shift configuration (DP-PSC), and the T/R-group phase-shift configuration (TR-PSC) strategies.
Furthermore, we derive the outage probabilities for the three proposed phase-shift configuration strategies
as well as for those of the
random phase-shift configuration and the independent phase-shift model, which constitute performance lower and upper bounds, respectively.
Then, the diversity order of each strategy is investigated based on the obtained analytical results. It is shown that 
the proposed DP-PSC strategy achieves
full diversity order simultaneously for users located on both sides of the STAR-RIS. Moreover, power scaling laws are derived for the three proposed strategies and for the random phase-shift configuration.
Numerical simulations 
reveal a performance gain if the users on both sides of the STAR-RIS are served by NOMA instead of OMA. Moreover, it is shown that the proposed DP-PSC strategy yields the same diversity order as achieved by STAR-RISs under the independent phase-shift model and a comparable power scaling law with only $4$ dB reduction in received power.

\end{abstract}

\begin{IEEEkeywords} 
Electromagnetics, hardware modelling, performance analysis, reconfigurable intelligent surfaces (RISs), simultaneous transmission and reflection.
\end{IEEEkeywords}
\section{Introduction}
With the rapid development of wireless communication technologies, Internet citizens have adapted to a lifestyle, where fast and reliable wireless connections are a fundamental component of their work and daily life. This new lifestyle imposes stringent requirements on the spectrum efficiency, latency, security, and coverage of wireless networks. Recently, the concept of smart radio environments~\cite{ahead} has been proposed based on a new and fundamental enabler, namely reconfigurable intelligent surfaces (RISs).
The key idea is that RISs can reconfigure the propagation of wireless signals and direct them to their intended receivers. This manipulation is facilitated by employing a large number of low-cost elements with reconfigurable electromagnetic (EM) responses. 
The low-cost and passive nature of RIS elements is one of the most important benefits compared to multi-antenna and active relay-aided wireless networks. As the passive elements do not require radio-frequency (RF) chains, RISs are scalable in size while achieving desirable power scaling and beamforming performances~\cite{renzo_diff}.

However, most of the existing research has investigated reflecting-only RISs~\cite{liu2020reconfigurable}, where each RIS element is only able to 
reflect signals to one side of the surface. As a result, the coverage of these RISs is greatly reduced and unnecessary topological constraints
have to be satisfied during their practical deployment. For example, transmitter and receiver have to be located on the same side of reflecting-only RISs~\cite{liu2020reconfigurable}.
As a remedy, the novel concept of simultaneous transmitting and reflecting RISs (STAR-RISs) was recently proposed~\cite{xu_star,liu_star}. Generally speaking, STAR-RISs can realize a full-space smart radio environment by simultaneously transmitting and reflecting the incident wireless signals onto both sides of the surface. To facilitate this, each STAR-RIS element has to dynamically impose two distinct tunable phase-shift values, one for the transmitted signal and one for the reflected signal~\cite{xu_star}. 
Given the above unique features, STAR-RISs have been regarded as a promising technology for future wireless networks, thus attracting extensive interest from both industry~\cite{doc,zhang2021intelligent,xu_vtmag} and academia~\cite{xu_star,mu2021simultaneously,star_coverage}.

\subsection{Prior Works}\label{prior}
Motivated by the various benefits of deploying smart surfaces in wireless networks, extensive research efforts have been devoted to the design and analysis of RIS-aided networks. In the following, we briefly summarize the recent research contributions on RISs and STAR-RISs.
\subsubsection{Optimization and Modeling of Conventional Reflecting-Only RISs}
In early 2019, the authors of~\cite{basar} demonstrated the benefits of RISs for wireless communications and separated the concept of RIS from other fields of study including meta-materials and optics. In~\cite{basar}, the basic hardware model of RISs was given, where each element was characterized by a phase-shift value. Based on this model, RIS-aided communication systems were investigated with the objective to improve energy efficiency~\cite{huang}, sum rate~\cite{mu2019exploiting}, and other performance metrics~\cite{sensing}. Furthermore, different practical hardware and channel models were studied. The authors of~\cite{rui_couple} proposed a practical phase-shift model, where the phase-shift and amplitude of each element are coupled. 
The authors of~\cite{9319694} proposed a mutual coupling-aware communication model for RIS which takes into account the mutual coupling correlation of adjacent RIS elements.
In~\cite{xu2020novel}, the authors proposed a novel channel model where the signals reflected by scatters and RIS elements are jointly considered as multipath components of the overall received envelope. The authors of \cite{vlc} and \cite{danufane2020path} investigated physics based channel models exploiting the Huygens-Fresnel principle and the Green's function, respectively. 
%

\subsubsection{Modeling and Analysis of STAR-RISs}
Compared to conventional reflecting-only RISs, the research on STAR-RISs is at a much earlier stage.
In~\cite{xu_star}, the authors proposed an independent phase-shift model for STAR-RIS and demonstrated the extended coverage of STAR-RIS compared to conventional RIS.
The benefits of STAR-RISs were further demonstrated in \cite{liu_star}, \cite{mu2021simultaneously}, and \cite{star_coverage}, where potential application scenarios, related optimization problems, and practical operating protocols were proposed.
Moreover, several STAR-RIS prototypes have been reported. In~\cite{doc}, NTT DOCOMO announced a successful trial of a transparent dynamic metasurface. Their implementation supports full reflection, full transmission (penetration) mode, and hybrid modes, which is similar to a STAR-RIS. In~\cite{zhang2021intelligent}, the authors presented a STAR-RIS prototype where the transmission and reflection coefficients of each element can be adjusted through positive-intrinsic-negative (PIN) diodes. In~\cite{xu_vtmag}, different categories of STAR-RIS hardware implementations, hardware models, and channel models were discussed and compared.
In~\cite{liu2021simultaneously}, the authors considered a coupled transmission and reflection phase-shift model for STAR-RISs. Based on this model, a power consumption minimization problem was formulated and solved using alternating optimization.

\subsection{Motivations and Contributions}

As outlined above, existing works on performance analysis and optimization of RIS-aided wireless networks mainly focused on conventional reflecting-only RISs. Since STAR-RISs introduce more adjustable parameters (i.e., transmission and reflection coefficients), these 
existing results are not applicable to
STAR-RIS-aided wireless networks.
Furthermore, the current works~\cite{xu_star,liu_star,mu2021simultaneously} on STAR-RISs assume that the transmission and reflection coefficients of each element can be independently adjusted. This, however, is non-trivial to realize in practice.
The reason is that for STAR-RISs to achieve independent control over both the transmitted and reflected signals, active and lossy STAR-RIS elements must be employed. However, this can only be achieved with metasurfaces with non-local effects~\cite{ahead} or patch-arrays with additional power supply, which would significantly increase manufacturing costs and reduce scalability.  
Therefore, it is important to consider purely passive STAR-RIS implementations and study their phase-shift limitations.
According to EM theory, for passive lossless STAR-RIS elements, the amplitudes and phases of the reflected and transmitted fields are constrained by boundary conditions and the law of energy conservation.
These theoretical constraints need to be considered in the modeling and the analysis of passive STAR-RISs.

To fully understand the limitations and to exploit the potentials of STAR-RISs, we propose a correlated\footnote{Note that the correlation investigated in this work is not due to mutual coupling, as studied in~\cite{9319694}. Here, we study the correlation between the transmission and reflection coefficients of a given STAR-RIS element.} transmission and reflection (T\&R) phase-shift model for passive-lossless STAR-RIS elements. Based on this model, we develop three practical phase-shift configuration (PSC) strategies for STAR-RIS-aided communication systems. For each strategy, the resulting communication performance is analyzed in terms of the outage probability and power scaling law to obtain useful insights.
The main contributions of this paper can be summarized as follows:
\begin{itemize}
    \item We propose a correlated T\&R phase-shift model for STAR-RISs,
    which is derived based on the equivalent surface current principle. To study the impact of the proposed phase-shift model on the performance of wireless systems, we investigate STAR-RIS-aided downlink communication systems employing orthogonal multiple access (OMA) and non-orthogonal multiple access (NOMA), respectively.
    \item We propose three practical PSC strategies for STAR-RISs with correlated phase shifts, namely the primary-secondary PSC (PS-PSC), diversity preserving PSC (DP-PSC), and T/R-group PSC (TR-PSC) strategies. We show that the PS-PSC strategy is suitable for generating channels with different channel gains for the users on different sides of the STAR-RIS, while the DP-PSC strategy produces channels with similar channel for all users.
    \item We derive the outage probability and power scaling law for each of the proposed PSC strategies.
    For comparison, 
    we consider STAR-RISs with an independent phase-shift model and random phase-shift configurations, respectively, as these two cases constitute performance upper and lower bounds, respectively. 
    We further deduce the diversity orders of the users for all proposed strategies and show that full diversity order can be simultaneously achieved for the users on both sides of the STAR-RIS with the DP-PSC strategy.
    \item We demonstrate the effectiveness of the proposed PSC strategies by comparing their radiation patterns. 
    Our numerical results reveal a performance gain of NOMA over OMA for the DP-PSC strategy.
    Furthermore, our simulations show that
    under the same channel conditions, 
    the DP-PSC strategy achieves a similar power scaling law as the performance upper bound with only $4$ dB degradation in received power.
\end{itemize}
Taking into account the T\&R phase-shift correlation, the preliminary study in~\cite{liu2021simultaneously} formulated optimization problems for STAR-RIS-aided wireless networks, which led to complicated solutions. These solutions entailed a high complexity and 
did not provide any insights for system design since a performance analysis was not feasible. Thus, in this paper, we pursue a different approach and analyzes the achievable diversity order and power scaling law of STAR-RIS-aided wireless systems for a correlated T\&R phase shift model, with the objective to provide valuable insights for system design.

\subsection{Organization}
The rest of the paper is organized as follows. In Section II, the proposed correlated T\&R phase-shift model for STAR-RISs is introduced, and the considered STAR-RIS-aided downlink transmission system is specified.
In Section III, different practical PSC strategies are proposed for STAR-RISs with correlated phase shifts.
In Section IV, the performance of different PSC strategies is analyzed and compared in terms of outage probability, diversity order, and power scaling law.
Numerical results are presented in Section V, and conclusions are drawn in Section VI.

\emph{Notations:} Scalars and vectors are denoted by lower/upper-case and bold-face letters, respectively. Rician distributed random variables are characterized by their shape parameter $K$ and scale parameter $\Omega$. $\mathbb{E}[x]$ and $\mathrm{Var}[x]$ denote the expected value and variance of a scalar random variable $x$. $|\xi|$ and $\angle{\xi}$ denotes the magnitude and the argument of a complex number, respectively. $f_{|h|}(x)$ denotes the probability density function (PDF) of a random variable $|h|$. 


\section{Correlated STAR-RIS Phase-Shift Model and System Model}\label{sec_hardware}
In this section, we present the proposed correlated T\&R phase-shift model for STAR-RIS. Then, we present the system model for the considered STAR-RIS-aided communication system.
However, we first introduce some prerequisites from physics to capture the characteristics of STAR-RIS hardware.

\begin{figure}[t!]
    \begin{center}
        \includegraphics[scale=0.45]{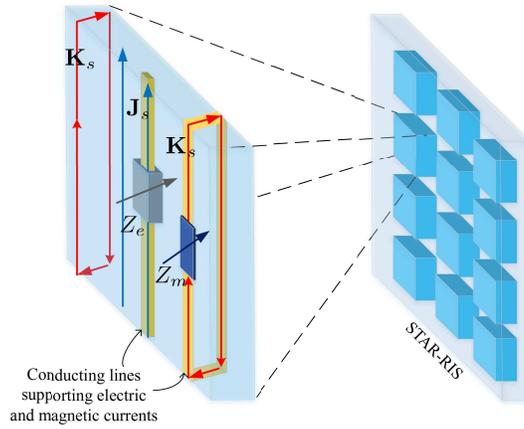}
        \caption{Conceptual illustration of the hardware model of the STAR-RIS.}
        \label{new_m}
    \end{center}
\end{figure}
\begin{figure}[t!]
    \begin{center}
        \includegraphics[scale=0.4]{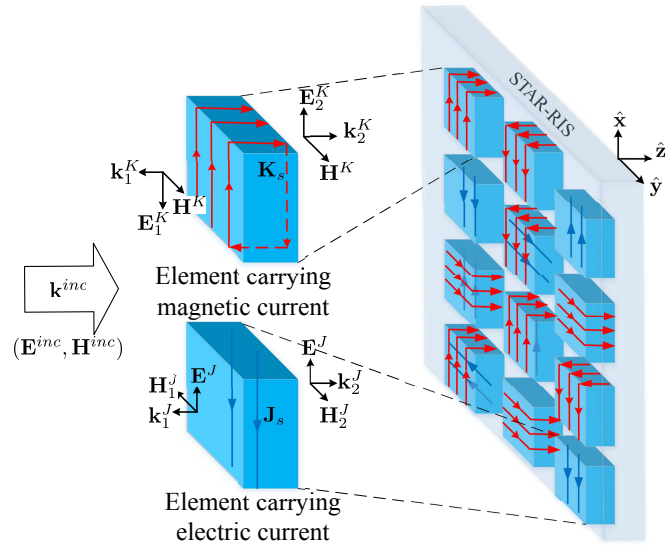}
        \caption{EM radiations of STAR-RIS elements carrying induced currents $\mathbf{J}_s$ and $\mathbf{K}_s$, where $\mathbf{k}^{inc}$, $\mathbf{k}^{J}_1$, $\mathbf{k}^{J}_2$, $\mathbf{k}^{K}_1$, and $\mathbf{k}^{K}_2$ denote the wave vector (wave number) of the corresponding waves.}
        \label{m2}
    \end{center}
\end{figure}

\subsection{Surface Equivalent Currents}
Consider a STAR-RIS whose elements are made of a single-layered passive material. Upon being illuminated by the incident EM wave, multiple currents are induced in each element, cf. Fig.~\ref{new_m}. For STAR-RIS modeling,
it is sufficient to consider only two types of currents\footnote{This is because in terms of radiation, the electric/magnetic charge layers and electric/magnetic current layers are equivalent~\cite{old}.}, namely the electric current and the magnetic current~\cite{old}. The electric current is generated by the flow of free electrical charges within the conducting part of the element media. Let $\mathbf{J}_s$ denote the electric current density. As shown in Fig.~\ref{m2}, since the electric field component of the impinging EM field is oscillating, $\mathbf{J}_s$ also oscillates and thus produces EM radiation with electric field $\mathbf{E}^J$, and magnetic fields $\mathbf{H}^J_1$ and $\mathbf{H}^J_2$ on the two sides of the STAR-RIS, respectively. The magnetic current density, $\mathbf{K}_s$, is generated by vortex (circular) currents within each STAR-RIS element~\cite{rothwell2018electromagnetics}. These currents are induced by the oscillating magnetic component of the EM field and they produce EM radiation with electric fields $\mathbf{E}^K_1$ and $\mathbf{E}^K_2$, and magnetic field $\mathbf{H}^K$. The strengths of these two types of currents are assumed to be proportional to the sum of the incident and radiated fields at the STAR-RIS.
Thus, the density of the electric and magnetic currents are respectively given by
\begin{align}\label{js}
    \mathbf{J}_s &= Z_e (\mathbf{E}^{inc}+\mathbf{E}^J),\\ \label{ks}
    \mathbf{K}_s &= Z_m (\mathbf{H}^{inc}+\mathbf{H}^K),
\end{align}
where $Z_e$ and $Z_m$ are the scalar electric and magnetic impedances of a particular element, respectively, and
$\mathbf{E}^{inc}$ and $\mathbf{H}^{inc}$ are the electric and magnetic components of the incident field, respectively.
To connect the EM fields on both sides of the STAR-RIS, we take into account the boundary conditions of the EM field as follows~\cite{old}:
\begin{align}\label{bound1}
    \mathbf{n}\times (\mathbf{H}^J_1-\mathbf{H}^J_2) = \mathbf{J}_s \text{\ and \ }
    \mathbf{n}\times (\mathbf{E}^K_1-\mathbf{E}^K_2) = \mathbf{K}_s,
\end{align}
where $\mathbf{n}$ is the unit vector perpendicular to the STAR-RIS.

\subsection{Correlated T\&R Phase-Shift Model}
The strengths of the transmitted and reflected signals are determined by the magnitudes of the electric components of the transmitted and reflected EM fields.
Assuming a vertically polarized wireless signal, the $\mathbf{E}$ fields are in $x$-direction and the $\mathbf{H}$ fields are in $y$-direction (see Fig.~\ref{m2}). Omitting the superscripts, we can rewrite these fields as
$\mathbf{E} = \hat{\mathbf{x}}E$ and $\mathbf{H} = \hat{\mathbf{y}}H$, where 
$\hat{\mathbf{x}}$ and $\hat{\mathbf{y}}$ are the unit vectors in $x$-direction and $y$-direction, respectively, and
$E$ and $H$ denote the complex amplitudes of the corresponding fields.
Specifically, considering an element with both induced electric and magnetic currents ($\mathbf{J}_s$ and $\mathbf{K}_s$), the transmission and reflection coefficients can be expressed as follows:
\begin{align}
T &= \beta^T\cdot e^{j\phi^T} = ({E}^J+{E}^K_2+{E}^{inc})/{E}^{inc},\\ \label{r}
    R &= \beta^R\cdot e^{j\phi^R} = ({E}^J+{E}^K_1)/{E}^{inc},
\end{align}
where $\beta^T$ and $\beta^R$ are the real-valued transmission and reflection amplitudes, respectively, and $\phi^T$ and $\phi^R$ are the corresponding phase-shift values for transmission and reflection.
Next, we study 
the correlation between $\phi^T$ and $\phi^R$ for a lossless STAR-RIS element.
According to EM theory, the change in energy of an EM field within an arbitrary volume $V$ is given as follows~\cite{rothwell2018electromagnetics}:
\begin{equation}\label{dw}
    \frac{dW}{dt} = -\int_{(\Sigma)} (\mathbf{E}\times\mathbf{H}) \ d\Sigma - \int_{(V)} \mathbf{J}\cdot \mathbf{E} \ dV,
\end{equation}
where $W$ is the EM energy within the volume $V$, $d/dt$ denotes the time derivative, $\mathbf{E}$ and $\mathbf{H}$ denote the EM fields at the boundary of $V$, $\mathbf{J}$ denotes the surface electric and magnetic current densities,
operators $\times$ and $\cdot$ denote the cross product and dot product between two vectors, respectively,
and $\Sigma$ denotes the closed surface of the chosen volume $V$. The second term in \eqref{dw} is the integration of the Poynting vector evaluated at the closed surface $\Sigma$, which reflects the energy loss due to EM radiation. The third term is a volume integration and it reflects the Ohmic heating and the power of other non-electrostatic forces within volume $V$. According to \eqref{dw}, if we choose $V$ as the volume of a STAR-RIS element, the second term represents the EM energy radiated towards the target users, while the third term represents the energy loss in the volume. Thus, a locally lossless element requires that the time-averaged value of $\mathbf{J}\cdot\mathbf{E}$ is equal to zero for the entire element. In the context of \eqref{js} and \eqref{ks}, this indicates that $Z_e$ and $Z_m$ should have purely imaginary values. 
By exploiting \eqref{js}-\eqref{dw}, we obtain the coupled phase-shift model for passive lossless STAR-RIS as given in the following proposition.

\begin{figure}[t!]
    \begin{center}
        \includegraphics[scale=0.35]{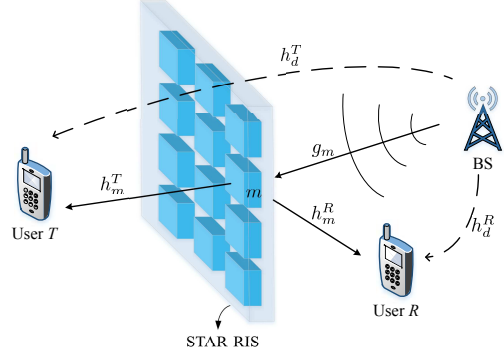}
        \caption{System model.}
        \label{sys1}
    \end{center}
\end{figure}

\begin{proposition}
(Proposed correlated T\&R phase-shift model)
For passive lossless STAR-RISs, the transmission and reflection amplitudes $\beta^T_m$ and $\beta^R_m$ and the phase shifts $\phi^T_m$ and $\phi^R_m$ of the $m$th element have to meet the following constraints:
\begin{equation}\label{amp}
    \beta^T_m = \sqrt{1-(\beta^R_m)^2},
\end{equation}
\begin{equation}\label{pha}
    \phi^R_m - \phi^T_m = \frac{\pi}{2} + \nu_m\pi, \ \nu_m=0\ \text{or}\ 1, \ \forall m=1,2,\cdots,M,
\end{equation}
where $M$ denotes the total number of STAR-RIS elements. $\nu_m$ is referred to as the \textit{auxiliary bit}\footnote{$\nu_m$ can only take on two values, 0 or 1. Thus, its value represents the two possible phase difference between $\phi^R_m$ and $\phi^T_m$, i.e., the phase difference is either $\pi/2$ or $3\pi/2$.} for the lossless STAR-RIS in the following
and provides
an additional degree of freedom which links the possible phase shift values between the transmission and reflection coefficients.

\begin{proof}
The relation in \eqref{amp} is a direct consequence of the law of energy conservation, i.e., the sum of the radiation power of both sides should be equal to the power of the incident signal~\cite{zhu2014dynamic}.
For \eqref{pha}, by exploiting \eqref{js}-\eqref{r}, it can be shown that the complex-valued T\&R coefficients of the passive STAR-RIS elements have to satisfy $|T+R|=1$ or $|T-R|=1$, or equivalently, $\beta^R_m\sqrt{1-(\beta^R_m)^2}\cos(\phi^R_m-\phi^T_m)=0$. Thus, for non-zero $\beta^R_m$, the phase difference must fulfill $\phi^R_m - \phi^T_m = \pi/2 \text{\ or\ } 3\pi/2$.
\end{proof}
\end{proposition}

\begin{remark}
The T\&R phase correlation in \eqref{pha} is a consequence of Maxwell's Equations and the boundary conditions. We note that independent phase shifts are achievable if the impedances $Z_e$ and $Z_m$ in \eqref{js} and \eqref{ks} can take on arbitrary complex values. However, a non-zero real part of $Z_e$ and $Z_m$ indicates that the STAR-RIS requires active or lossy elements, which might significantly increase the manufacturing cost.
\end{remark}

\subsection{System Model}\label{net_model}
To investigate the impact of the proposed correlated T\&R phase-shift model given in \eqref{pha} on wireless system performance, we consider a STAR-RIS-aided two-user downlink system, as illustrated in Fig.~\ref{sys1}. A single-antenna base station (BS) serves two single-antenna users, which are located on different sides of the STAR-RIS. The user located in the transmission region of the STAR-RIS is referred to as user T, while the user located in the reflection region is referred to as user R.
The channel between the BS and the $m$th STAR-RIS element is denoted by $g_m$ and the channels between the $m$th STAR-RIS element and user T and user R are denoted by $h^T_m$ and $h^R_m$, respectively. 
Let $h^{T/R}_d$ denote the direct link between the BS and user T/R.
All channels are assumed to be independent and identically distributed (i.i.d.) Rician fading channels\footnote{According to~\cite{book}, i.i.d. fading can be achieved by employing uniform linear arrays with $\lambda/2$-separation between adjacent antennas.}. Thus, for $\chi \in \{T,R\}$, indicating user T/R, $h^\chi_m$ can be expressed as follows:
\begin{align}\label{rice1}
    h^\chi_m \!&=\! \sqrt{\frac{\rho^\chi_0}{(d^\chi_m)^{\alpha}}}\Big( \sqrt{\frac{K^{\chi}_h}{K^{\chi}_h\!+\!1}}h^{\rm{LoS},\chi}_m \!+\! \sqrt{\frac{1}{K^\chi_h\!+\!1}}h^{\rm{NLoS},\chi}_m \Big),
\end{align}
where $d^\chi_m$ denotes the distance between the $m$th STAR-RIS element and user $\chi$, $\alpha$ is the path loss exponent, $K^\chi_h$ denotes the Rician factor, $\rho^\chi_0$ is the path loss at a reference distance of 1 meter, $h^{\rm{LoS},\chi}_m$ is the LoS component, and $h^{\rm{NLoS},\chi}_m$ is the non-line-of-sight (NLoS) component which is Rayleigh fading.
Channels $g_m$ and $h^\chi_d$ are modeled similarly to \eqref{rice1} with Rician factors $K_g$ and $K^\chi_d$, respectively.

The performance of STAR-RIS-aided networks depends on the choice of the phase shifts ($\phi^T_m$, $\phi^R_m$) and the amplitudes ($\beta^T_m$, $\beta^R_m$) applied by each element.
This is because the end-to-end channel between the BS and user $\chi$ is given by:
\begin{equation}
    H^\chi = \sum_{m=1}^M g_mh^\chi_m\beta^\chi_m e^{j\phi^\chi_m}+ h^\chi_d, \ \chi \in \{T,R\}.
\end{equation}
For ease of presentation, we rewrite the overall channel by separating the amplitude terms and the phase-shift terms as follows:
\begin{equation}\label{sep}
     H^\chi =  \sum_{m=1}^M \beta^\chi_m |g_m||h^\chi_m| \exp\{ j(\angle{g_m}+\angle{h^\chi_m+\phi^\chi_m}) \}+h^\chi_d,
\end{equation}
where $\angle{\xi}$ denotes the argument (complex angle) of a complex number $\xi$. 

\subsection{Multiple Access}
In this paper, both OMA and NOMA are considered for the BS to serve the users. In OMA, the BS serves both users in orthogonal frequency bands of equal size. In NOMA, the BS sends a superimposed signal to both users in the same time/frequency resource block.
Let 
$s^\chi$ and $c_\chi$ denote the information symbol and the power allocation coefficient for user $\chi$, respectively.
In the following, we outline both considered multiple access schemes.
\subsubsection{OMA}
In this case, the BS communicates with both users in orthogonal frequency bands employing frequency-division multiple access (FDMA).
Thus, the achievable rate for user $\chi$ is given by:
\begin{equation}\label{rate_oma}
    R^{\text{OMA}}_\chi = \frac{1}{2}\log_2\Big( 1+\frac{P_{BS}c^2_\chi|H^\chi|^2}{\sigma^2_0/2} \Big),
\end{equation}
where $P_{BS}$ and $\sigma^2_0$ are the transmit power of the BS and the variance of the additive white Gaussian noise (AWGN) at both users, respectively.
In \eqref{rate_oma}, we assume that the users employ orthogonal frequency bands of equal size. Thus, the bandwidth and the noise are both reduced by $1/2$.

\subsubsection{NOMA}
For NOMA, successive interference cancellation (SIC) is employed at the user enjoying better channel conditions.
For simplicity, we assume that user R has the better channel, i.e., $|H^R|\ge|H^T|$. To ensure that SIC is successfully carried out and to guarantee user fairness, the power allocation coefficients follow $c_R<c_T$. Therefore, user R will first detect the signal of user T via SIC, before detecting its own signal. The achievable rate for user R to
detect the message of user T is given by
\begin{equation}
    R^{\text{NOMA}}_{R,T} = \log_2\Big(1+ \frac{P_{BS}c^2_T|H^R|^2}{P_{BS}c^2_R|H^R|^2+\sigma^2_0} \Big).
\end{equation}
Then, user R can detect its own signal after subtracting the signal of user T via SIC. Hence, the achievable rate of user R is given as follows:
\begin{equation}\label{gamma_rr}
    R^{\text{NOMA}}_{R,R} = \log_2\Big(1+ \frac{P_{BS}c^2_R|H^R|^2}{\sigma^2_0}\Big).
\end{equation}
In contrast, user T will directly detect its own signal by treating the signal of user R as interference. The corresponding achievable rate is given by:
\begin{equation}\label{gamma_tt}
     R^{\text{NOMA}}_{T,T} = \log_2\Big(1+ \frac{P_{BS}c^2_T|H^T|^2}{P_{BS}c^2_R|H^T|^2+\sigma^2_0}\Big).
\end{equation}

\section{Practical Phase-Shift Configuration Strategies for STAR-RISs}\label{sec_design}
In this section, we propose three STAR-RIS PSC strategies for the proposed correlated phase-shift model in \eqref{pha}. 
We note that due to the T\&R phase-shift correlation in \eqref{pha}, existing results on phase-shift optimization for conventional RISs
and STAR-RISs with independent phase shifts
are not applicable for the considered passive STAR-RISs.

\subsection{Primary-Secondary Phase-Shift Configuration (PS-PSC) Strategy}
For the PS-PSC strategy, 
we adjust the phases of the T\&R coefficients while
assuming that the power ratio between the amplitudes of the T\&R coefficients is fixed to the same value for each STAR-RIS element.
The STAR-RIS can be optimized for enhancing the channel gain of one user, referred to as the primary user, while simultaneously serving the other user, referred to as the secondary user.
Based on this strategy, the PSC of the STAR-RIS elements can be determined in the following two steps.
Without loss of generality, assume that user R is the primary user. In the first step, the phase terms $\{\phi^R_1,\cdots,\phi^R_M\}$ are configured to maximize the channel gain of user R without considering the phase-shift correlation constraint in \eqref{pha}. This yields the following optimization problem:
\begin{equation}\label{out_pri}
   \max_{\phi^R_1,\cdots,\phi^R_M} |H^R|^2, \ \text{s.t.} \ \phi^R_m \in [0,2\pi), \ \forall m=1,2,\cdots,M.
\end{equation}

In the second step, the $\{\phi^T_1,\cdots,\phi^T_M\}$ are optimized to maximize the overall channel gain of user T for given $\{\phi^{R^*}_1,\cdots,\phi^{R^*}_M\}$ taking into account the  phase-shift correlation model in \eqref{pha}. In this case, the optimization of ${\phi_m^{T}}$ under constraint \eqref{pha} is equivalent to optimizing the auxiliary bits $\{\nu_1,\cdots,\nu_M\}$ of the STAR-RIS, which leads to the following optimization problem:
\begin{equation}\label{out_sec}
 \max_{\nu_1,\cdots,\nu_M}|H^T|^2, \ \text{s.t.} \ \nu_m = 0\ \text{or}\ 1, \ \forall m=1,2,\cdots,M.
\end{equation}
Since constraint \eqref{pha} is not taken into account for the maximization with respect to $\phi^R_1,\cdots,\phi^R_M$ in \eqref{out_pri}, the optimal values of both $\phi^R_1,\cdots,\phi^R_M$ and $\nu_1,\cdots,\nu_M$ can be obtained in closed form if the channel state information is known.
According to the cophase condition~\cite{liu2020reconfigurable}, the optimal solution to problem \eqref{out_pri} is given by:
\begin{equation} \label{pri_phi}
    \phi^{R^*}_m = (\angle{h^R_d}-\angle{h^R_m}-\angle{g_m}) \mod \ 2\pi.
\end{equation}

For the secondary user, the phase shift according to the cophase condition would follow as $\phi^{T'}_m =  (\angle{h^T_d}-\angle{h^T_m}-\angle{g_m}) \mod \ 2\pi$. However, this phase shift value might not be possible because of \eqref{pha} for the given $\phi^{R^*}_m$.
Thus, we obtain the optimal choice of the auxiliary bits as follows:
\begin{equation}\label{nu}
    \nu^*_m = 
    \begin{cases}
      0 & \text{if $\phi^T_0 \leq \phi^{R^*}_m < \phi^T_0 + \pi$,}\\
      1  & \text{if}\  \phi^T_0 - \pi \leq \phi^{R^*}_m < \phi^T_0,
    \end{cases}       
\end{equation}
where $\phi^T_0= \angle{h^T_d}$ is the phase of the direct link between user T and the BS. 
The PSC in \eqref{nu} effectively selects the $\phi^T_m$ value which is closer to $\phi^{T'}_m$ given the two possible choices, i.e., $\phi^{T}_m = \phi^{R^*}_m-\pi/2$ and $\phi^{T}_m = \phi^{R^*}_m-3\pi/2$.

\begin{table*}[!t]
\centering
\begin{tabular}{|c|c|c|c|c|}
\hline
($\Delta\phi^R_m-\Delta\phi^T_m)\in$  & $[0,\pi)$ & $[\pi,2\pi)$ & $[-\pi,0)$ & $(-2\pi,-\pi)$ \\ \hline
$\phi^{T^*}_m$ & $\phi'_m -\frac{\pi}{4}$   & $\phi'_m -\frac{3\pi}{4}$   & $\phi'_m +\frac{\pi}{4}$   & $\phi'_m +\frac{3\pi}{4}$   \\ \hline
$\nu^*$  & 0 & 1 & 1 & 0 \\ \hline
$\phi^{R^*}_m$ & $\phi'_m +\frac{\pi}{4}$   &  $\phi'_m +\frac{3\pi}{4}$  & $\phi'_m -\frac{\pi}{4}$   &  $\phi'_m -\frac{3\pi}{4}$  \\ \hline
\end{tabular}
\caption{Diversity preserving phase-shift configuration strategy, $ \phi'_m = (\Delta\phi^R_m+\Delta\phi^T_m)/2$.} \label{table_rules}
\end{table*}

\subsection{Diversity Preserving Phase-Shift Configuration (DP-PSC) Strategy}
One of the claimed benefits of STAR-RISs is that they can achieve full diversity orders for both the user T and the user R~\cite{xu_star}. 
Here, we show that this is possible even for STAR-RISs with correlated T\&R phase shifts. To this end, we propose a DP-PSC strategy.
For given amplitude coefficients, the phase shifts and the values of the auxiliary bits are determined as follows: 
First, according to the cophase condition, we calculate the phase-shift value that would maximize the channel gains of both users without taking into account the phase-shift correlation constraint. This leads to
\begin{equation}\label{d-p_phase}
    \Delta\phi^\chi_m = (\angle{h^\chi_d}-\angle{h^\chi_m}-\angle{g_m})\mod \ 2\pi.
\end{equation}
Then, taking the phase correlation into account, the PSC for the DP-PSC strategy is determined based on the calculated phase-shift values $\Delta\phi^R_m$ and $\Delta\phi^T_m$, 
and is given in Table~\ref{table_rules} along with the corresponding auxiliary bit $\nu$.
To illustrate the idea behind this PSC, let $\delta^\chi_m = |\phi^{\chi^*}_m-\Delta\phi^\chi_m|$ denote the phase error for user $\chi$.
The PSC given in Table~\ref{table_rules} imposes the identical phase errors on both users, i.e., $\delta^R_m=\delta^T_m$. Moreover, it can be verified that $\delta^\chi_m \leq \pi/4$. This property is crucial for the DP-PSC strategy to exploit the spatial diversity of the STAR-RIS, cf. Section~\ref{sec_performance}.

\subsection{T/R-Group Phase-Shift Configuration (TR-PSC) Strategy}
Similar to the mode-switching protocol in~\cite{liu_star}, for the proposed TR-PSC strategy, the elements of the STAR-RIS are partitioned into two groups. The elements in the T group employ full transmission and serve only the user T, while
the elements in the R group employ full reflection and serve only the user R.
For the fully reflecting elements, we have $\beta^R_m=1$ and $\beta^T_m=0$, and for the fully transmitting elements, we have $\beta^T_m=1$ and $\beta^R_m=0$.
Due to the decoupling of transmission and reflection at the element level, the phase correlation constraint in \eqref{pha} does not apply to the TR-PSC strategy.
Let $\mathcal{T}$ denote the set of indices of the elements in the T group.
Then, the transmission and reflection coefficients for the TR-PSC strategy are given by:
\begin{align}\label{mode_phi_1}
    \phi^R_m = \angle{h^R_d}-\angle{h^R_m}-\angle{g_m},  \ m \notin \mathcal{T}, \\ \label{mode_phi_2}
    \phi^T_m = \angle{h^T_d}-\angle{h^T_m}-\angle{g_m}, \ m \in \mathcal{T}.
\end{align}

\begin{figure}[t!]
    \begin{center}
        \includegraphics[scale=0.35]{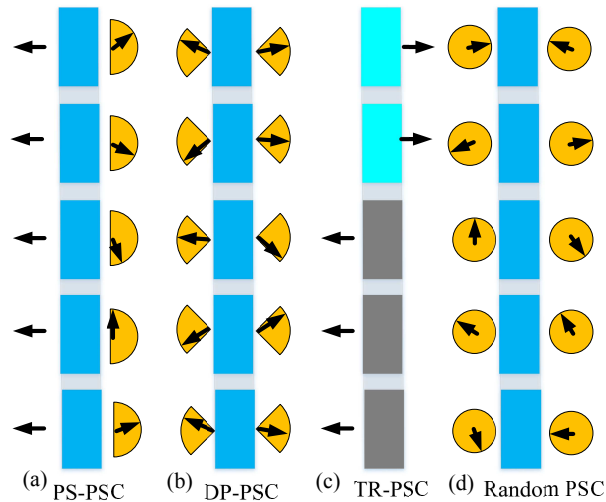}
        \caption{Conceptual illustration of different STAR-RIS PSC strategies.}
        \label{designs}
    \end{center}
\end{figure}

\subsection{Summery and Discussion}
In Fig.~\ref{designs}, we illustrate the three proposed PSC strategies for STAR-RISs and the random PSC as a baseline scheme. According to \eqref{sep}, at user $\chi$, the overall phase of the signal received from element $m$ is $\angle{g_m}+\angle{h^\chi_m}+\phi^\chi_m$. The small arrows in Fig.~\ref{designs} indicate these overall phase terms.
Under the assumption of i.i.d. fading channels, the yellow areas denote the 
possible ranges of the values of the overall phase terms.
Note that these sectors are different in size for the different users and for different designs, and indicate the possible deviation of the actual phase shift from the optimal values given by the cophase condition. 
As shown in Fig.~\ref{designs}(a), the overall phase terms for the primary user under the PS-PSC strategy and for both users under the TR-PSC strategy are equal to the values obtained from the cophase condition.
In contrast, the yellow sectors of the secondary user for the PS-PSC strategy have central angles of $180$ degrees. The central angles of the yellow sectors of users under the DP-PSC strategy and the random PSC are $90$ and $360$ degrees, respectively.
These deviations of the overall phase terms are different for different PSC strategies, which leads to significant performance difference for the users.

\section{Performance Analysis}
\label{sec_performance}

In the following, the communication performance of the T and R users is analyzed for the proposed PSC strategies and performance lower and upper bounds are established.
Specifically, we first derive the outage probabilities and diversity orders for the proposed PSC strategies.
Then, we study the power scaling law, which quantifies how the average received power scales with the number of STAR-RIS elements. 

\subsection{Outage Probability}

\subsubsection{OMA}

For FDMA, 
the BS serves the users in two orthogonal frequency bands of equal
sizes employing frequency-flat STAR-RIS transmission and reflection coefficients. 
Suppose that for user $\chi$ the target data rate is $\Tilde{R}_\chi$. Then, according to \eqref{rate_oma}, the outage probability for the user is given by
\begin{equation}\label{out_oma}
    P^{\text{OMA}}_{\text{out},\chi}  =\mathrm{Pr}\big\{{R}^{\text{OMA}}_\chi< \Tilde{R}_\chi\big\} = \mathrm{Pr}\Big\{ |H^\chi|^2 < \frac{(2^{2\Tilde{R}_\chi}-1)\sigma^2_0}{2c^2_\chi P_{BS}} \Big\},
\end{equation}
where $\mathrm{Pr}\{A\}$ denotes the probability of event $A$.
For ease of exposition, we define the following term as the OMA channel gain threshold for target data rate $\Tilde{R}_\chi$:
\begin{equation}
    \tau^{\text{OMA}}_\chi (\Tilde{R}_\chi)\triangleq  (2^{2\Tilde{R}_\chi}-1)\sigma^2_0/(2c^2_\chi).
\end{equation}

\subsubsection{NOMA}
Suppose that for NOMA user $\chi$, the target data rate is $\Tilde{R}_\chi$. Moreover, we consider the case where user R has the better channel condition.
For given target data rates $\Tilde{R}_T$ and $\Tilde{R}_R$
of the users, their outage probabilities can be formulated as follows:
\begin{align}\label{out,r}
\begin{split}
    P^{\text{NOMA}}_{\text{out},R} = \mathrm{Pr}\Big\{ |H^R|^2 < 
    \max &\big\{\frac{(2^{\Tilde{R}_T}-1)\sigma^2_0}{\big[c^2_T-(2^{\Tilde{R}_T}-1)c^2_R\big]P_{BS}}\\
    &, \frac{(2^{\Tilde{R}_R}-1)\sigma^2_0}{c^2_RP_{BS}}  \big\}
    \Big\},
\end{split}
\end{align}

\begin{equation}\label{out,t}
    P^{\text{NOMA}}_{\text{out},T} = \mathrm{Pr}\Big\{ |H^T|^2 < \frac{(2^{\Tilde{R}_T}-1)\sigma^2_0}{\big[c^2_T-(2^{\Tilde{R}_T}-1)c^2_R\big]P_{BS}} \Big\}. 
    \end{equation}

For convenience, we define the following two NOMA channel gain thresholds:
\begin{align}
    \tau^{\text{NOMA}}_R &\triangleq \max \big\{\frac{(2^{\Tilde{R}_T}-1)\sigma^2_0}{c^2_T-(2^{\Tilde{R}_T}-1)c^2_R}, \frac{(2^{\Tilde{R}_R}-1)\sigma^2_0}{c^2_R} \big\},\\
    \tau^{\text{NOMA}}_T &\triangleq \frac{(2^{\Tilde{R}_T}-1)\sigma^2_0}{c^2_T-(2^{\Tilde{R}_T}-1)c^2_R}.
\end{align}
Since both $\Tilde{R}_R$ and $\Tilde{R}_T$ appear in \eqref{out,r}, for NOMA, user R will experience an outage if the target data rate of user T is chosen too large. This is because in SIC, user R will declare an outage either when it cannot decode its own message or when it cannot decode the message of user T.

\subsection{PS-PSC Strategy}
For the the primary user (user R in the considered case), the overall channel gain in \eqref{sep} can be rewritten as follows:
\begin{equation}\label{hr}
    H^R = \beta^R\sum_{m=1}^M |g_m| |h^R_m| + h^R_d,
\end{equation}
where $\beta^R$ is the amplitude of the reflection coefficient shared by all STAR-RIS elements, i.e., $\beta^R_m = \beta^R$, $\forall m=1,2,\cdots,M$, in \eqref{amp}.
\subsubsection{Outage Probability}
Given the overall channel gain in \eqref{hr}, we obtain the following result:
\begin{theorem}\label{t_pri}
For the primary user (user R), the asymptotic outage probability for NOMA and OMA can be expressed as follows:
\begin{align}\label{out1}
\begin{split}
P^R_{out}(\tau_R)&=\frac{2^{M+1}(K^R_h+1)^M(K_g+1)^M(K_d+1)}{(2M+2)!\ \Omega_h^{M}\Omega^M_g\Omega^\chi_d}
(\beta^R)^{-2M}\\ &\cdot e^{-MK^R_h-MK_g-K^R_d}(\tau_R)^{M+1} P_{BS}^{-(M+1)},
\end{split}
\end{align}
where $\tau_R = \tau^{\text{NOMA}}_R$ for NOMA and  $\tau_R = \tau^{\text{OMA}}_R$ for OMA, and $\Omega^\chi_d$ is the scale parameter of the direct link of user $\chi$.
\begin{proof}
See Appendix~A.
\end{proof}
\end{theorem}
Next, we investigate the diversity order of user $\chi$, which is defined as follows~\cite{liu_enhancing}:
\begin{equation}\label{d_define}
d^{\chi} = -\lim_{P_{BS} \to \infty}\frac{\log P^{\chi}_{out}(\tau_\chi)}{\log P_{BS}}.
\end{equation}
\begin{corollary}
The diversity orders of the primary user for the PS-PSC strategy are identical for both NOMA and OMA. Exploiting \eqref{out1} and using \eqref{d_define}, the diversity order for the primary user is given by:
\begin{equation}\label{d_pri}
    d_{\text{primary}} = M+1.
\end{equation}
\begin{proof}
Combining \eqref{out1} with \eqref{d_define}, it is straightforward to obtain \eqref{d_pri}.
\end{proof}
\end{corollary}

For the secondary user, the overall channel can be rewritten as follows:
\begin{align}\label{h_what}
    H^T &= \sum_{m=1}^M |g_m| |h^T_m|e^{j\phi'_m}\beta^T e^{j\phi^T_m} + h^T_d,\\ \label{ht}
    &= \beta^T\sum_{m=1}^M |g_m| |h^T_m| e^{j(\phi^T_0+\delta_m)}+ h^T_d,
\end{align}
where $e^{j\phi'_m}$ is the phase of $g_mh^T_m$, $\delta_m = \phi'_m + \phi^T_m - \phi^T_0$ is the residual phase, which satisfies $-\pi/2<\delta_m < \pi/2$ according to \eqref{pha} and \eqref{nu}, $\phi^T_0$ is the phase of the direct channel between user T and the BS,
and $\beta^T$ is the amplitude of the transmission coefficient which is assumed to be equal for all STAR-RIS elements, i.e., $\beta^T_m = \beta^T = \sqrt{1-(\beta^R)^2}$, $\forall m=1,2,\cdots,M$.
As a result, for the secondary user, the STAR-RIS elements can be regarded as a one-bit phase shifter. The asymptotic behaviour and diversity order for one-bit phase shifters has been studied in~\cite{onebit}. Hence, given the presence of the direct BS-user link, the diversity order of the secondary user (user T) for both NOMA and OMA is given by\footnote{The diversity order of user R and user T switches if user T is the primary user.}:
\begin{equation}\label{d_sec}
    d_{secondary} = \frac{M+3}{2}.
\end{equation}
\begin{remark}
The diversity orders given in \eqref{d_pri} and \eqref{d_sec} include the contribution of the direct channel $h^\chi_d$. For the scenario where the direct link is blocked, it can be shown that the diversity orders decrease by one, which leads to the following results:
\begin{equation}\label{diversity_prime}
d'_{\text{primary}} = M,\  d'_{\text{secondary}} = \frac{M+1}{2}.
\end{equation}
\end{remark}

\subsubsection{Power Scaling Law}

Due to the asymmetric PSC, the power scaling laws will be different for user R and user T. 
For convenience, let $\mu^\chi_h$ and $\sigma^\chi_h$ denote the expected value and the variance of the amplitude of $h^\chi_m$, respectively. The following results hold~\cite{simon2002probability}
\begin{align}
\mu^\chi_{h}&=\frac{1}{2}\sqrt{\frac{\pi\Omega^\chi_{h}}{(K^\chi_{h}+1)}}L_{1/2}(-K^\chi_{h}),\\
\sigma^\chi_h &= \Omega^\chi_h-(\mu^\chi_h)^2,
\end{align}
where $L_{1/2}(x)$ denotes the Laguerre polynomial.
For the primary user, according to \eqref{hr} and given that the $|h^R_m|$ are assumed to be i.i.d. Rician random variables, the expected value and variance of $H^R$ can be express as follows:
\begin{align}\label{ehr}
\begin{split}
    \mathbb{E}[|H^R|] &= |g|\beta^R\sum_m \mathbb{E}[|h^R_m|] + \mathbb{E}[|h^R_d|]\\&
    =M\mu^R_h|g|\beta^R + \mathbb{E}[|h^R_d|],
\end{split}
\end{align}
\begin{equation}\label{varhr}
    \mathrm{Var}[|H^R|] = M\sigma^R_h|g|^2(\beta^R)^2 + \mathrm{Var}[|h^R_d|],
\end{equation}
where $|g|=\mathbb{E}[|g_m|]$ is the expected magnitude of the channel between the $m$th STAR-RIS element and the BS.

For the secondary user, according to \eqref{ht}, the overall channel $H^T$ is a complex-valued sum of the channel amplitude terms $|h^T_m|$ and the phase-shift terms $e^{j(\phi^T_0+\delta_m)}$. Due to the complexity of this mathematical expression, we first present a lemma as a building block for the power scaling analysis for the secondary user. 
We rewrite the overall channel of the secondary user as $H^T=h^T_s+h^T_d$, where $h^T_s = \beta^T\sum_{m=1}^M |g_m| |h^T_m| e^{j(\phi^T_0+\delta_m)}$ is the STAR-RIS-aided channel.

\begin{lemma}\label{lemma_1}
The PDF of the magnitude of $h^T_s$ can be approximated as follows\footnote{Numerical results presented in \cite{xu2020novel} showed that the Kullback–Leibler (KL) divergence between the approximated PDF in \eqref{xu_rician} and the exact distribution of $|h^T_s|$ is less than 0.05 for a 16$\times$16 element RIS.}:
      \begin{equation}\label{xu_rician}
          f_{|h^T_s|}(x)=\frac{x}{\beta^2}e^{-\frac{x^2+\alpha^2}{2\beta^2}}I_0\Big(\frac{x\alpha}{\beta}\Big),
      \end{equation}
      where
      \begin{align}\label{shape_factors}
          \alpha = 2M\mu^T_h /\pi,\ 
          \beta^2 = \frac{M}{2}\Omega^T_h.
      \end{align}
      
\begin{proof}
See Appendix B.
\end{proof}
\end{lemma}

\begin{theorem}\label{theorem_pri_scaling}
Under the PS-PSC strategy, the power scaling laws of the two users are given by:
\begin{align}\label{scaling1}
\begin{split}
    P_{\text{primary}} = P^R_r 
    =& M^2\cdot (\mu^R_h|g|\beta^R)^2 +
    M\cdot\big\{\sigma^R_h|g|^2(\beta^R)^2 \\ &+ 2\mu^R_h|g|\beta^R \mathbb{E}[|h^R_d|] \big\}+\mathbb{E}[|h^R_d|^2],
    \end{split}
\end{align}
and
\begin{align}
\begin{split}\label{secondary_power}
    P_{\text{secondary}}=P^T_r
    =& \frac{4}{\pi^2} M^2\cdot (\mu^T_h|g|\beta^T)^2 + M\cdot\big\{\sigma^R_h|g|^2(\beta^T)^2\\
    &+ 2\mu^T_h|g|\beta^T \mathbb{E}[|h^T_d|] \big\}+\mathbb{E}[|h^T_d|^2].
    \end{split}
\end{align}
\begin{proof}
For the primary user, the received power can be calculated as follows:
\begin{equation}\label{proofpr}
P^R_r = \mathbb{E}[|H^R|^2] = \mathrm{Var}[|H^R|] + (\mathbb{E}[|H^R|])^2.
\end{equation}
Thus, the power scaling law of the primary user can be obtained by substituting \eqref{ehr} and \eqref{varhr} into \eqref{proofpr}. For the secondary user, we have $P_r(T)= \int_0^\infty x^2  f_{|h^T_s|}(x) dx+ \mathbb{E}[|h^T_d|^2]$. By using the the PDF of $|h^T_s|$ in \eqref{xu_rician}, the scaling law in \eqref{secondary_power} is proved.
\end{proof}
\end{theorem}

\begin{remark}
The power scaling law in \eqref{scaling1} can be further simplified for STAR-RISs with sufficiently large numbers of elements. For the case where $M\gg 1$, the received power simply scales with $M^2$ in the large $M$ regime, i.e.,
\begin{equation}\label{remark_pr}
    P^\chi_r \propto \Tilde{f}_\chi\cdot M^2,
\end{equation}
where
$\Tilde{f}_R=(\mu^R_h\beta^R)^2$ and $\Tilde{f}_T=(\frac{2}{\pi}\mu^T_h\beta^T)^2$.
This result is consistent with the existing power scaling analysis for conventional reflecting-only RIS~\cite{basar}. However, the received power of the secondary user is reduced by a factor of $(2/\pi)^2$ compared with that of the primary user.
\end{remark}

\subsection{DP-PSC Strategy}
In this subsection, we demonstrate that the DP-PSC strategy indeed achieves full diversity order for the users on both sides of the STAR-RIS. 
For the convenience of the subsequent analysis, we rewrite the phase-shift value of the $m$th element for user $\chi$ as follows:
\begin{equation}\label{phi_chi_di}
    \phi^\chi_m = \Delta\phi^\chi_m + \delta^\chi_m,
\end{equation}
where $\Delta\phi^\chi_m$ is the phase shift that achieves cophasing for user $\chi$ and $\delta^\chi_m = \phi^\chi_m-  (\angle{h^\chi_d}-\angle{h^\chi_m}-\angle{g_m})\mod \ 2\pi$ is the deviation from this optimal value due to the phase correlation.
Based on the channel model presented in Section~\ref{sec_design}, by plugging \eqref{phi_chi_di} into \eqref{sep}, the overall channel for user $\chi$ can be rewritten as follows:
\begin{equation}\label{h_delta}
    H^\chi = \beta^\chi \sum_{m=1}^M |g_m||h^\chi_m| \exp\{ j(\angle{h^\chi_d}+\delta^\chi_m) \}+h^\chi_d.
\end{equation}

\subsubsection{Outage Probability}
Due to the complexity of the DP-PSC strategy, it is very challenging to obtain the exact channel distribution for both users. To address this issue, we introduce the following theorem based on which the diversity orders of user T and user R can be obtained.

\begin{theorem}
The outage probabilities of user T and user R are upper-bounded by:
\begin{align}\label{out_bound1}
\begin{split}
    P^\chi_{out}(\tau_\chi) &\leq \big(F(\tau_\chi,K_g,K^\chi_h,\Omega_g,\Omega^\chi_h)\big)^M \\&
    \cdot\Big(1-Q(\sqrt{2K^\chi_d},\sqrt{2\tau_\chi(K^\chi_d+1)/\Omega^\chi_d P_{BS}})  \Big),
\end{split}
\end{align}
where $\tau_\chi = \tau^{\text{NOMA}}_\chi$ for NOMA and $\tau^{\text{OMA}}_\chi$ for OMA, $Q(a,b)$ is the Marcum Q-function, which is defined as follows:
\begin{equation}\label{mq_noma}
  Q(a,b)=\int_{b}^{\infty}x \cdot \exp\big(-\frac{x^2+a^2}{2}\big)I_0(ax)dx,
\end{equation}
$I_0(x)$ denotes the modified Bessel function of the first kind, and $F(\tau_\chi,K_g,K^\chi_h,\Omega_g,\Omega^\chi_h) =\mathrm{Pr}\big\{|g_m||h^\chi_m|\leq \sqrt{\frac{\tau_\chi}{P_{BS}}} \big\}$. A closed-form expression for $F$ can be obtained using the exact PDF of $|g_m||h^\chi_m|$, i.e., a product of two Rician variables~\cite{simon2002probability}, which is omitted here for brevity.

\begin{proof}
In \eqref{h_delta}, the phase difference between any two terms in the summation is always less than $\pi$, i.e., $|\delta^\chi_p-\delta^\chi_q|\leq \pi, \ \forall p,q\in\{1,2,\cdots,M\}$.
Thus, according to the law of cosines, we have the following proposition:
\begin{align}
\begin{split}
&|H^\chi|<\sqrt{\frac{\tau_\chi}{P_{BS}}}\  \Rightarrow \ \Big(|h^\chi_d|<\sqrt{\frac{\tau_\chi}{P_{BS}}}\Big)\land \\ &
\Big(|g_1||h^\chi_1|<\sqrt{\frac{\tau_\chi}{P_{BS}}}\Big)\land\cdots \land\Big(|g_M||h^\chi_M|<\sqrt{\frac{\tau_\chi}{P_{BS}}}\Big),
\end{split}
\end{align}
where $\mathrm{P} \Rightarrow \mathrm{Q}$ means proposition $\mathrm{P}$ leads to $\mathrm{Q}$ and $\land$ is the \textit{logical AND operator}. Since $|H_1|, |H_2|, \cdots, |H_M|$ are i.i.d. random variables, according to probability theory, we have the following result:
\begin{align}\label{out_bound}
\begin{split}
    P^\chi_{out}\!\leq\! \Big(\mathrm{Pr}\big\{|g_m||h^\chi_m|\!\leq\! \sqrt{\frac{\tau_\chi}{P_{BS}}} \big\}\Big)^M 
    \cdot\mathrm{Pr}\big\{ |h^\chi_d|\!\leq\! \sqrt{\frac{\tau_\chi}{P_{BS}}} \big\}.
\end{split}
\end{align}
Finally, the second term on the right-hand side of \eqref{out_bound} can be calculated using the Rician PDF of $|h^\chi_d|$ as follows:
\begin{align}
\begin{split}
    \mathrm{Pr}\big\{ |h^\chi_d|\leq & \sqrt{\frac{\tau_\chi}{P_{BS}}}  \big\} 
    =1-Q\big(\sqrt{2K^\chi_d},\sqrt{\frac{2\tau_\chi(K^\chi_d+1)}{\Omega^\chi_d P_{BS}}}\big).
    \end{split}
\end{align}
This completes the proof.
\end{proof}
\end{theorem}
\begin{corollary}
For the DP-PSC strategy, full diversity order is achieved for both user R and user T, i.e.,
\begin{equation}\label{d_pre}
    d^R=d^T=M+1.
\end{equation}
\begin{proof}
The right-hand side of \eqref{out_bound} is a multiplication of $M+1$ terms. According to the system model, $|h^\chi_d|$ is Rician distributed with
shape prameter $K_d$ and scale parameter $\Omega_d = \mathbb{E}[|h^\chi_d|^2]$.
Since in the high transmit power regime, i.e., $P_{BS}\to \infty$, 
the outage probability is determined by the PDF of $|h^\chi_d|$ near zero, we consider the Taylor expansion of the PDF near the origin: $f_{|h^\chi_d|}(x) = \frac{2(K^\chi_d+1)}{\Omega_d}x + o(x)$. Therefore, we have the following result:
\begin{align}
    \mathrm{Pr}\big\{|h^\chi_d|\!\leq\! \sqrt{\frac{\tau_\chi}{P_{BS}}} \big\}
    &\!=\! \int_0^{\sqrt{\frac{\tau_\chi}{P_{BS}}}} f_{|h^\chi_d|}(x)dx 
    \!\approx\! \frac{2}{\Omega_h}(K^\chi_d\!+\!1) \frac{\tau_\chi}{P_{BS}}.
\end{align}
The remaining terms in \eqref{out_bound}, i.e., $\mathrm{Pr}\big\{ |g_m||h^\chi_m|\leq \sqrt{\frac{\tau_\chi}{P_{BS}}}\big\}$ can be evaluated in a similar manner. Thus, we obtain for $P^\chi_{out}(\tau_\chi)$ the following expression:
\begin{equation}
    P^\chi_{out}(\tau_\chi) \sim (P_{BS})^{-(M+1)}.
\end{equation}
Based on the definition of the diversity order given in \eqref{d_define}, we conclude that the full diversity order $M+1$ is achieved for both users.
\end{proof}
\end{corollary}

\subsubsection{Power Scaling Law}
As can be observed from Fig.~\ref{designs}(b), the DP-PSC strategy is symmetric for the two users. As a result, similar power scaling laws apply for both users. For simplicity, we consider the scenario, where $M$ is large and the power carried by the direct channel is negligible. For the end-to-end channel given in \eqref{h_delta}, we obtain the following theorem.

\begin{theorem}\label{t_d_scaling}
For the DP-PSC strategy, the power scaling laws for both users can be expressed as follows:
\begin{equation}\label{scaling_d_pre}
    P^\chi_r \propto \frac{8}{\pi^2}M^2 (\mu^\chi_h|g|\beta^\chi)^2 + \Big(1-\frac{2}{\pi}\Big)M\sigma^\chi_h|g|^2(\beta^\chi)^2.
\end{equation}
\begin{proof}
The received power can be expressed as $ P^\chi_r = \mathbb{E}[|H^\chi|^2]$.
Furthermore, by substituting $\Delta^\chi=\pi/2$ in the proof process given in Appendix B, the approximate PDF of $|H^\chi|$ is obtained as:
\begin{equation}\label{xu_rician2}
          f_{|H^\chi|}(x)=\frac{x}{\beta^2}e^{-\frac{x^2+\alpha^2}{2\beta^2}}I_0\Big(\frac{x\alpha}{\beta}\Big),
      \end{equation}
      where
      \begin{align}\label{shape_factors2}
          \alpha = 2\sqrt{2}M\mu^\chi_h /\pi,\ 
          \beta^2 = (1-2/\pi)\frac{M}{2}\Omega^\chi_h.
      \end{align}
Finally, by integrating the PDF obtained in \eqref{xu_rician2}, the proof is completed.
\end{proof}
\end{theorem}

\begin{remark}
Similar to the PS-PSC strategy, the average received power under the DP-PSC strategy also scales with $M^2$ in the large $M$ regime. However, under the same channel conditions and using equal power splitting, i.e., $\beta^R=\beta^T$, comparing the average power received by the primary user and secondary user for the PS-PSC strategy, $P^{\text{primary}}_{\text{PS-PSC}}$ and $P^{\text{secondary}}_{\text{PS-PSC}}$, respectively, and the average power received by the users for the DP-PSC strategy, $P_{\text{DP-PSC}}$, we have:
\begin{equation}
    P^{\text{primary}}_{\text{PS-PSC}} > P_{\text{DP-PSC}} > P^{\text{secondary}}_{\text{PS-PSC}}.
\end{equation}
\end{remark}

\subsection{TR-PSC Strategy}
For the TR-PSC strategy, the STAR-RIS is equivalent to a composite smart surface containing a reflecting-only RIS with $M_R$ elements and a transmitting-only RIS with $M_T$ elements, where $M_R+M_T = M$. As a result, for each reflecting-only/transmitting-only element, the phase shift of the reflection/transmission coefficient can be configured without taking into account the correlation in \eqref{pha}.
As cophasing can be achieved on both sides, the end-to-end channel for user $\chi$ is given by
\begin{equation}\label{out_mode}
    H^\chi = \sum_{m'=1}^{M_\chi}|g_{m'}||h^\chi_{m'}|+h^\chi_d.
\end{equation}

\subsubsection{Outage Probability}
Considering \eqref{out_mode}, the end-to-end channel is the sum of the amplitudes of the cascaded channels $|g_{m'}||h^\chi_{m'}|$, which are perfectly aligned in phase. Thus, this expression is in a similar form as the end-to-end channel of the primary user for the PS-PSC strategy.
\begin{theorem}
For the TR-PSC strategy, the asymptotic outage probability for both users can be expressed as follows:
\begin{align}\label{out_best}
\begin{split}
P^\chi_{out}(\tau_\chi)&=\frac{2^{M_\chi+1}(K^R_h+1)^{M_\chi}(K^R_d+1)}{(2M_\chi+2)! \ (\Omega^R_h)^{M_\chi}\Omega^R_d}
(\beta^R)^{-M_\chi/2}\\ &
\cdot e^{-M_\chi K_s-K_d}(\tau_\chi)^{M_\chi+1} P_{BS}^{-(M_\chi+1)},
\end{split}
\end{align}
where $\tau_\chi = \tau^{\text{NOMA}}_\chi$ for NOMA and $\tau_\chi = \tau^{\text{OMA}}_\chi$ for OMA.
\begin{proof}
The proof of this theorem is similar to that of \textbf{Theorem~\ref{t_pri}}, and is omitted here for brevity.
\end{proof}
\end{theorem}
According to \eqref{out_best}, for the TR-PSC strategy, the diversity orders are given by
\begin{equation}\label{d_best}
    d^R= M_R+1 \text{\ \ and\ \ } d^T = M_T+1.
\end{equation}

\subsubsection{Power Scaling Law}
By comparing the PSC of the TR-PSC strategy in \eqref{mode_phi_1} and \eqref{mode_phi_2} with the configuration in \eqref{pri_phi}, we observe that the power scaling laws of the TR-PSC strategy can be deduced from \textbf{Theorem \ref{theorem_pri_scaling}}.
\begin{corollary}\label{coro_mode}
For the TR-PSC strategy, the power scaling law for both users can be expressed as follows:
\begin{align}\label{scaling_mode}
\begin{split}
    P^\chi_r &\propto M_\chi^2\cdot (\mu^\chi_h|g|\beta^\chi)^2 \\ & 
    + M_\chi\cdot\big\{\sigma^\chi_h|g|^2(\beta^\chi)^2 + 2\mu^\chi_h|g|\beta^\chi \mathbb{E}[|h^R_d|] \big\}+\mathbb{E}[|h^R_d|^2].
    \end{split}
\end{align}
\begin{proof}
The proof of this corollary is similar to the proof of \textbf{Theorem \ref{theorem_pri_scaling}}, and is omitted for brevity.
\end{proof}
\end{corollary}
\begin{remark}
\textbf{Corollary~\ref{coro_mode}} provides an important insight regarding the differences between the power scaling laws for the DP-PSC and TR-PSC strategies.
For a STAR-RIS with a total of $M$ elements, in the large $M$ regime, the received power for DP-PSC scales with $M^2$ on both sides of the surface while TR-PSC can only produce powers that scale with respectively $M^2_R$ and $M^2_T$ on the two sides of the STAR-RIS. Furthermore, since $M_R, M_T \leq M$, the power differences between the two strategies are more pronounced for large numbers of elements.
\end{remark}

\subsection{Performance Lower and Upper Bounds}\label{lower_upper}
To have performance baselines for the proposed PSC strategies for STAR-RIS-aided communication systems, we proposed lower and upper bounds.

For the performance lower bound, we assume that the STAR-RIS employs a random PSC strategy. For the performance upper bound, the STAR-RIS is assumed to be able to perform independent phase-shift adjustments for both the transmitted and reflected signals, as was done in~\cite{xu_star,liu_star,mu2021simultaneously}.
The end-to-end channel gain and the achievable rate of the primary user for the PS-PSC strategy achieve this performance upper bound.
This is because for the primary user, the STAR-RIS phase-shift coefficients are chosen without considering the phase correlation constraint. 
Thus, by designating user R or user T as the primary user in \eqref{d_pri} and \eqref{scaling1}, the diversity order and power scaling law of the proposed performance upper bound is obtained.
Consequently, in the following, we focus on the analysis of the proposed performance lower bound.

\subsubsection{Outage Probability}

For the random PSC strategy, we assume that for all $m$, the phase-shifts of the reflection coefficients are randomly chosen, i.e., $\phi^R_m$ is uniformly distributed within $[0,2\pi)$. The overall end-to-end channel between the BS and user $\chi$ is given by
\begin{equation}\label{sep2}
     H^\chi = |g|\beta^\chi \sum_{m=1}^M |h^\chi_m| \exp\{ j(\angle{g_m}+\angle{h^\chi_m+\phi^\chi_m}) \}+ h^\chi_d.
\end{equation}
Moreover, due to the randomness of the STAR-RIS phase shifts, $\phi^\chi_m$, the entire phase term in \eqref{sep2} can be regarded as a uniformly distributed random variable, which leads to the following theorem. For convenience, we denote the first term in \eqref{sep2} as $h^\chi_s$, and thus the overall channel is given by $H^\chi = h^\chi_s+ h^\chi_d$.

\begin{theorem}
For the random PSC strategy, if the number of STAR-RIS elements is large, $|h^\chi_s|$ can be approximated by the following PDF:

\begin{equation}\label{random_pdf}
f_{|h^\chi_s|}(x)=\frac{2x}{\Omega^\chi_r}e^{-\frac{x^2}{\Omega^\chi_r}},
\end{equation}
where $\Omega^\chi_r=M|g|^2(\beta^\chi)^2\Omega^\chi_h$. 
\begin{proof}
We consider the real (inphase) and imaginary (quadratic) parts of $h^\chi_s$:
\begin{align}
    &T_c = |g|\beta^\chi\sum|h^\chi_m|\cos(\angle{g_m}+\angle{h^\chi_m+\phi^\chi_m}),\\
    &T_s = |g|\beta^\chi\sum|h^\chi_m|\sin(\angle{g_m}+\angle{h^\chi_m+\phi^\chi_m}).
\end{align}
According to the central limit theorem (CLT), for large $M$, the distributions of $T_c$ and $T_s$ can be approximated by Gaussian distributions
with zero means and variances $\mathbb{E}[T^2_c]=\mathbb{E}[T^2_s]=M|g|^2(\beta^\chi)^2\Omega^\chi_h/2$. Thus, $|h^\chi_s|=\sqrt{T^2_c+T^2_s}$ follows a Rayleigh distribution with the PDF given in \eqref{random_pdf}.
\end{proof}
\end{theorem}
For the considered system, the outage probability of user $\chi$ can be analyzed as follows:
\begin{align}\label{out_define}
\begin{split}
    &P^\chi_{out}(\gamma^\chi_0) = \mathrm{Pr}\Big \{ \frac{(|h^\chi_s|^2 + |h^\chi_d|^2) c^2_\chi P_{BS}}{\sigma_0^2} < \gamma^\chi_0 \Big \} \\
    & \leq \mathrm{Pr}\Big\{|h^\chi_s|^2< \frac{\tau_\chi}{P_{BS}} \Big\}\cdot \mathrm{Pr}\Big\{|h^\chi_d|^2< \frac{\tau_\chi}{ P_{BS}} \Big\}\\
    &=\Big (1- e^{-\frac{\tau_\chi}{\Omega^\chi_r P_{BS}}}\Big )\cdot \Big(1-Q(\sqrt{2K^\chi_d},\sqrt{\frac{2\tau_\chi(K^\chi_d+1)}{\Omega^\chi_d P_{BS}}})  \Big),
    \end{split}
\end{align}
Thus, for the random PSC strategy, the diversity order of both users can be calculated using \eqref{d_define} and \eqref{out_define}:
\begin{equation}\label{d_random}
    d_{\text{random}}^R=d_{\text{random}}^T = 2.
\end{equation}

\subsubsection{Power Scaling Law}
Considering the PDF of $h^\chi_s$ given in \eqref{random_pdf}, we have $\mathbb{E}[|h^\chi_s|] = \Omega^\chi_r=M|g|^2(\beta^\chi)^2\Omega^\chi_h$. Then, the power scaling law for the random PSC strategy is given by
\begin{equation}\label{eq_scaling_random}
    P^\chi_r \propto \mathbb{E}[|h^\chi_s|^2]+\mathbb{E}[|h^\chi_d|^2] = M|g|^2(\beta^\chi)^2\Omega^\chi_h + \mathbb{E}[|h^\chi_d|^2].
\end{equation}
Note that the result in \eqref{eq_scaling_random} scales linearly with $M$. Thus, compared to the three proposed PSC strategies, the average power received for the users increase significantly slower with $M$ for the random PSC strategy.

\begin{table*}[t!]
\small
\centering
\begin{tabular}{|c|c|c|c|c|}
\hline
STAR-RIS phase-shift strategy  & PS-PSC & DP-PSC & TR-PSC & Random PSC \\ \hline
Diversity order of user R         & $M+1$      & $M+1$ & $M_R+1$   & $2$   \\ \hline
Diversity order of user T         & $(M+3)/2$  & $M+1$ & $M_T+1$  & $2$    \\ \hline

Power scaling law       & $\propto M^2$  & $\propto M^2$ & $\propto M^2_\chi$    & $\propto M$    \\ \hline

\end{tabular}
\caption{Comparing different PSC strategies.}
\label{table_diversity}
\end{table*}

\subsection{Summary and Discussion}
The main results of this section are summarized in
Table~\ref{table_diversity}\footnote{In this table, user R is assumed to be the primary user for the PS-PSC strategy.}. It can be observed
that 
the PS-PSC strategy is suitable for creating asymmetric channel gains while the diversity preserving strategy is suitable for balancing the performances of both users.
Moreover, Table~\ref{table_diversity} reveals that the DP-PSC strategy provides the best performance in terms of diversity gain and power scaling.

\begin{figure*}[t!]
\centering
\subfigure[PS-PSC]{\label{n0}
\includegraphics[width= 1.5in]{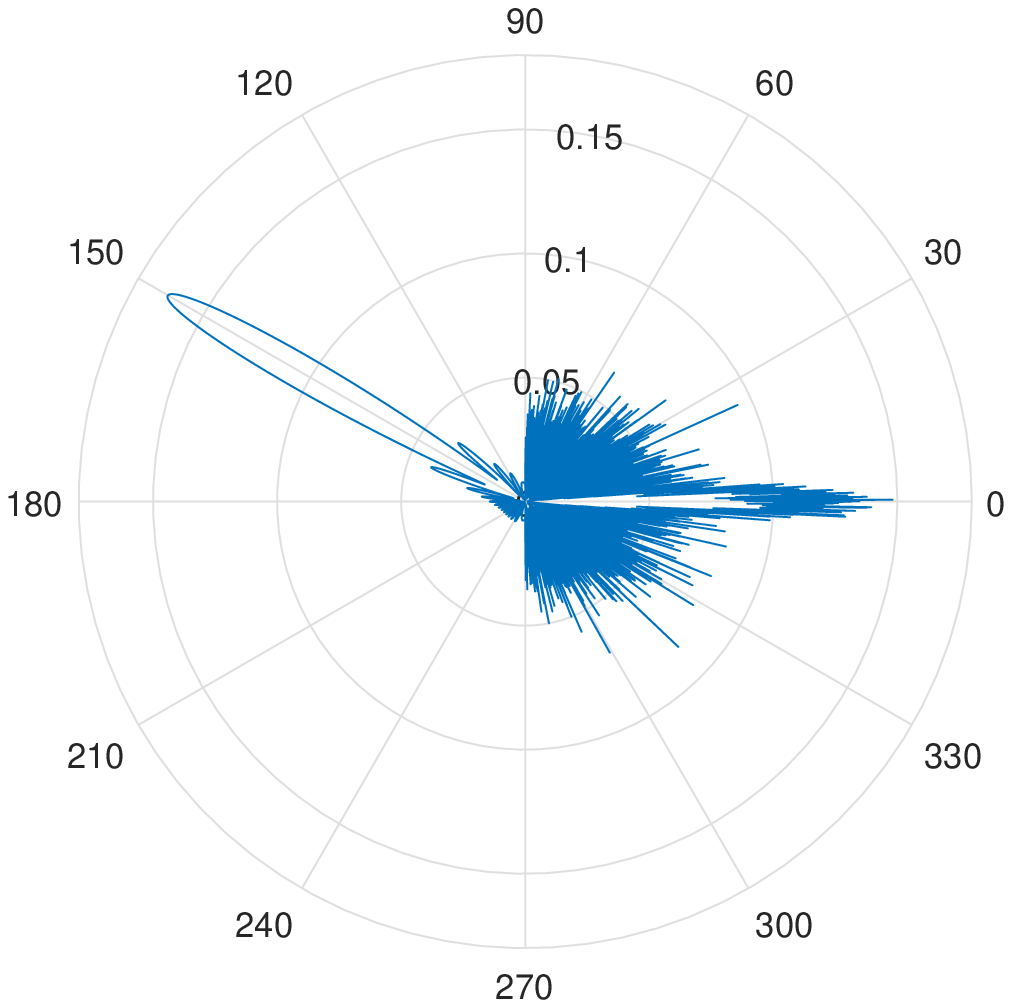}}
\subfigure[DP-PSC]{\label{na}
\includegraphics[width= 1.5in]{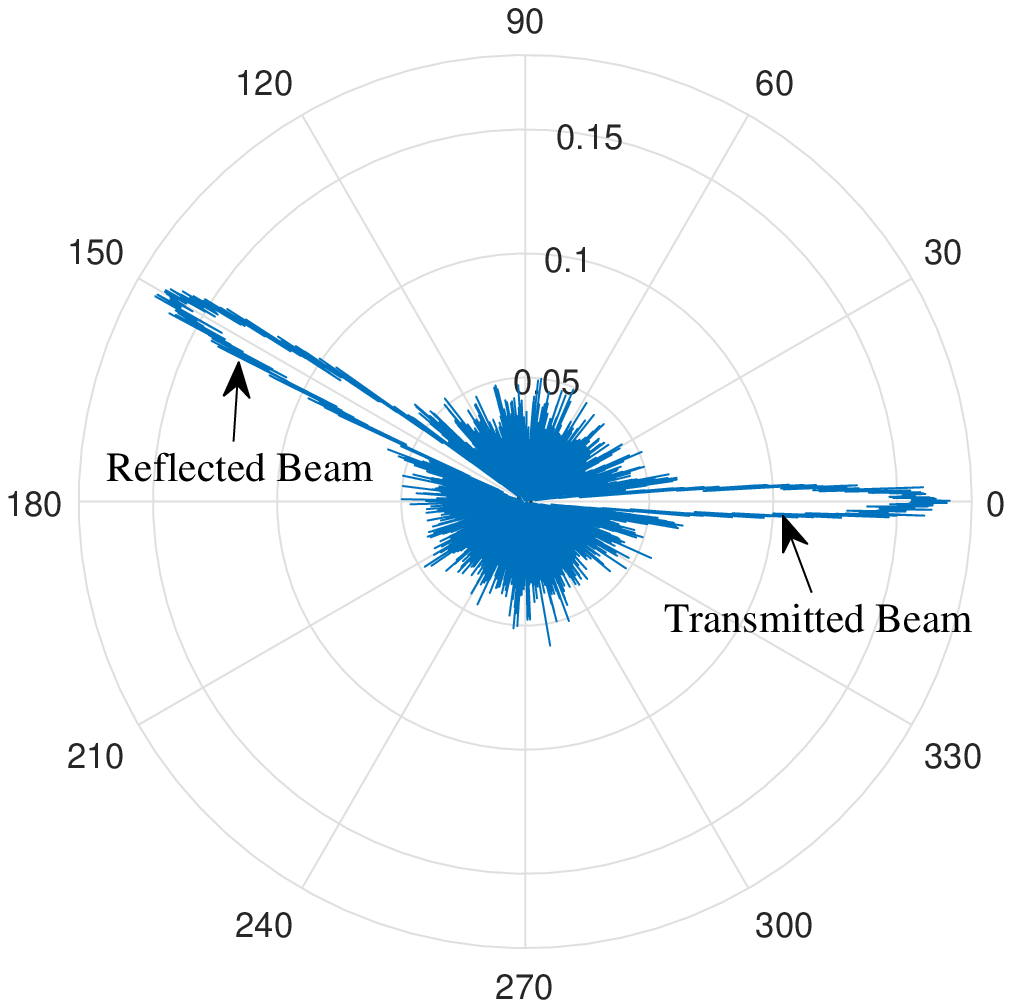}}
\subfigure[TR-PSC]{\label{nd}
\includegraphics[width= 1.5in]{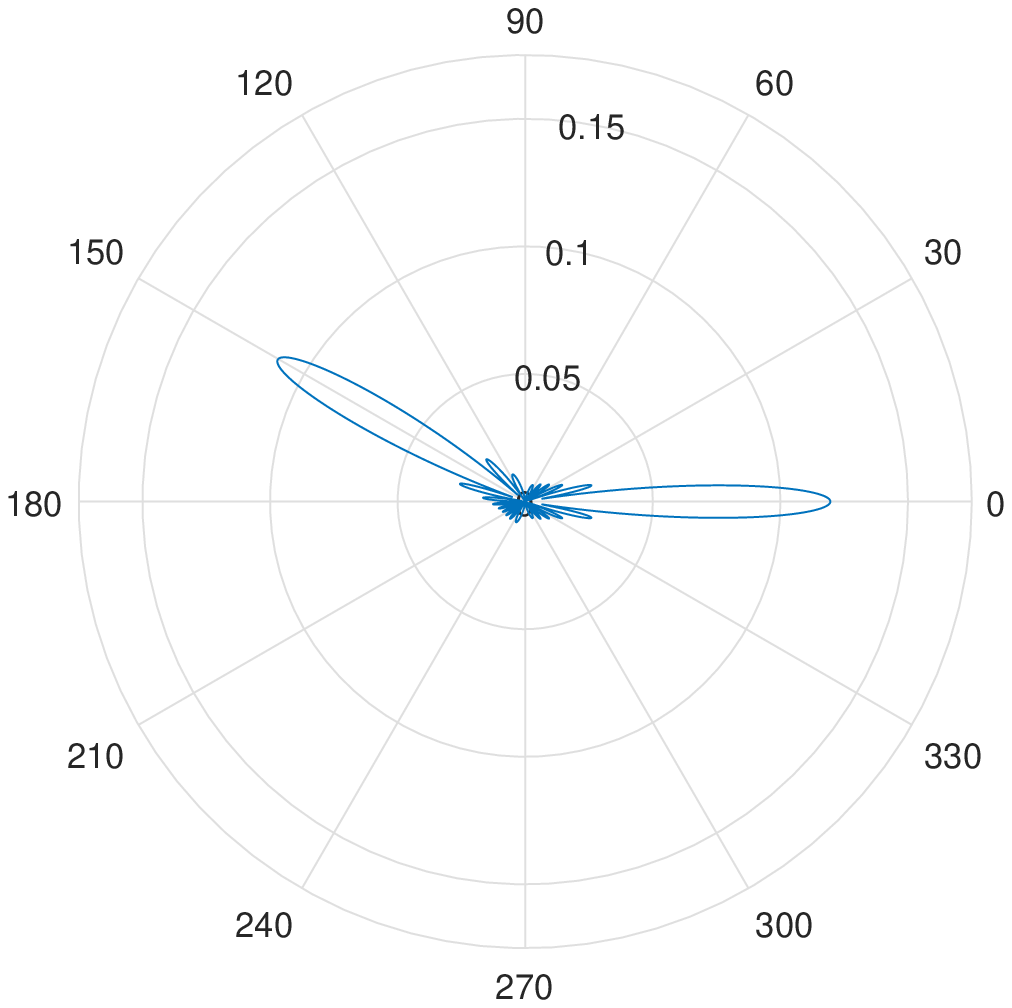}}
\subfigure[Random PSC]{\label{nb}
\includegraphics[width= 1.5in]{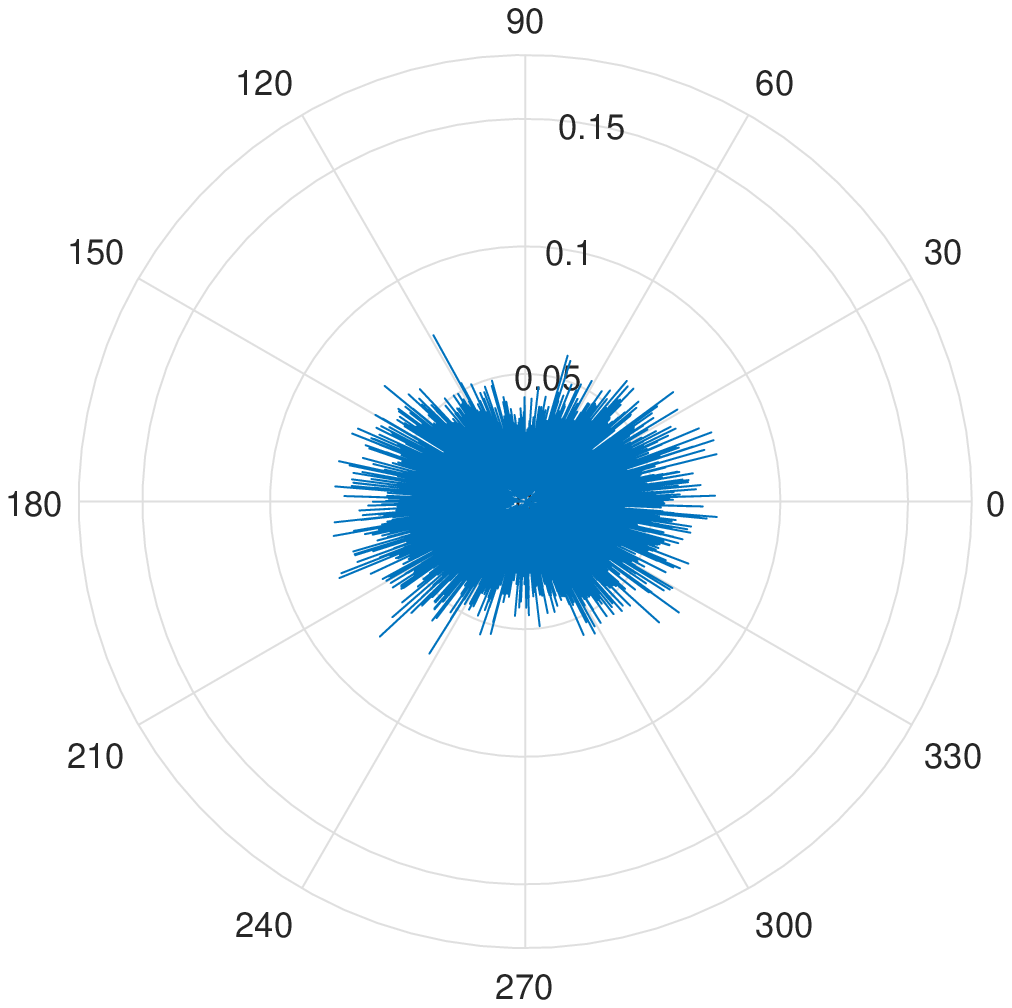}}
\caption{Radiation patterns of STAR-RIS for different PSC strategies.}\label{nice}
\end{figure*}

\section{Numerical Results}\label{num}

In this section, simulation results are provided to verify the performance of the proposed PSC strategies for STAR-RISs. For our simulations, we assume that the STAR-RIS is a uniform planar array consisting of $M=M_h\times M_v$ elements, where $M_h$ and $M_v$ denote the number of elements along the horizontal and vertical directions, respectively. The spacing between adjacent elements is half of the carrier wavelength. 
We assume that two users are located at angular positions of $0^\circ$ and $150^\circ$ with equal distances of $10$ m from the STAR-RIS, while the BS is located at an angular position of $270^\circ$, with a distance of $50$ m from the STAR-RIS. 
All STAR-RIS-user channels are modeled as Rician fading channels with path loss exponent $\alpha = 2.2$ and Rician factor $K=1.3$ dB. The noise power for both users is set to $\sigma^2_0 = -50$ dBm.
For the PS-PSC and DP-PSC strategies, the amplitudes of the T\&R coefficients of all elements are set to $\beta^T=\beta^R=1/\sqrt{2}$.
For the TR-PSC strategy, we assume that $M_R=M_T=M/2$. For the random PSC strategy, both the T\&R phase shifts of each element are randomly generated within $[0,2\pi)$.

\subsection{Beam Patterns of Different PSC Strategies}\label{num_1}
In Fig.~\ref{nice}, we compare the angular patterns of the proposed PSC strategies
and the baseline random PSC strategy. Here, we consider a STAR-RIS with 
18$\times$18 elements.
As can be seen from Fig.~\ref{n0}, for the PS-PSC strategy, the reflected beam for the primary user (i.e., user R in the direction of 150$^\circ$) has a 
larger power gain than the transmitted beam for the secondary user. 
Moreover, at the transmission side of the STAR-RIS (angular direction ranges of $(270^\circ,360^\circ)$ and $[0^\circ,90^\circ)$), considerable side lobes are caused by power leakage. This is expected since for the PS-PSC strategy, the phase shifts of the elements cannot be perfectly aligned for the secondary user, i.e., user T. In Fig.~\ref{na}, for the DP-PSC strategy, the reflected and transmitted beams exhibit similar power levels. However, different from the reflected beam of the primary user for the PS-PSC strategy in Fig.~\ref{n0}, the two beams for the PS-PSC strategy have noisy edges. 
In Fig.~\ref{nd}, both the reflected and transmitted beams for the TR-PSC strategy exhibit only small side lobes. However, compared to the DP-PSC strategy, the beam gain of the TR-PSC strategy is smaller since only part of elements are operated in the full reflection/transmission mode, which reduces the spatial diversity, see \eqref{d_best}.
The results shown in Fig.~\ref{n0}-(c) highlight an interesting trade-off between the power of the desired signal and the power of undesired interference.
Specifically,
although the beam gain for the DP-PSC strategy is larger than that for the TR-PSC strategy, DP-PSC also imposes more power leakage in non-intended directions leading to significant interference. For the TR-PSC strategy, the beam gain in the target direction is sacrificed but almost no interference in other directions is caused. Based on this observation, 
selecting the optimal PSC strategy for specific application scenarios constitutes an interesting topic for further investigation. 
For the baseline random PSC strategy, in  Fig.~\ref{nb}, no dedicated target beams can be observed, and the signal power is distributed over the entire space. The above results confirm the importance of employing suitable PSC strategies for STAR-RISs.

\begin{figure}[t!]
    \begin{center}
        \includegraphics[scale=0.6]{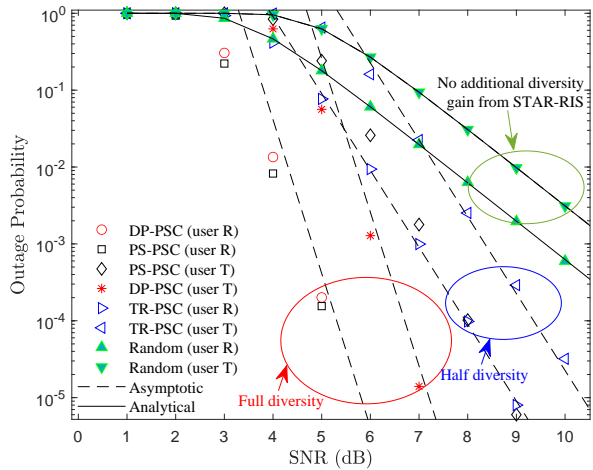}
        \caption{Outage probabilities for OMA users employing different PSC strategies.}
        \label{compare_out}
    \end{center}
\end{figure}

\subsection{Outage Probability and Diversity Order}
For the following simulation results, the transmit SNR ($P_{BS}/\sigma^2_0$) was varied between $0$ dB and $12$ dB.
We consider an asymmetric channel by setting the expected power of the T\&R channels as $\mathbb{E}[|h^R_m|^2] = 2 \mathbb{E}[|h^T_m|^2]$.
In Fig.~\ref{compare_out}, we present simulation results for the outage probabilities of both users for the proposed PSC strategies and the baseline random PSC strategy. 
To verify our performance analysis, asymptotic results are presented for the proposed PSC strategies exploiting \eqref{out1} and \eqref{out_best}, while analytical results are presented for the random PSC strategy exploiting \eqref{out_define}. 

As can be observed, the primary user for the PS-PSC strategy achieves the lowest outage probability and full diversity order, since 
the STAR-RIS can align the reflected channels
with the direct channel for the primary user. However, due to the correlated T\&R phase-shift model, the secondary user suffers a noticeable performance loss as the available choices for the transmission phase shifts are limited. Moreover, it can be observed that the primary user and the secondary user achieve full diversity order and reduced diversity order, respectively, which is consistent with \eqref{d_pri} and \eqref{d_sec}. 
For the DP-PSC strategy, as can be seen from Fig.~\ref{compare_out}, the outage probabilities of both users exhibit similar slopes and can achieve full diversity order, which is consistent with \eqref{d_pre}. 
For the TR-PSC strategy, since only half of the STAR-RIS elements are used for transmission/reflection (i.e., reduced array gain), there is a considerable performance loss for both users and only half of the maximum diversity order can be achieved, as is expected from \eqref{d_best}. Furthermore, the random PSC strategy yields the worst performance and achieves no additional diversity gain, which is in accordance with \eqref{d_random}.
\begin{figure}[h!]
    \begin{center}
        \includegraphics[scale=0.6]{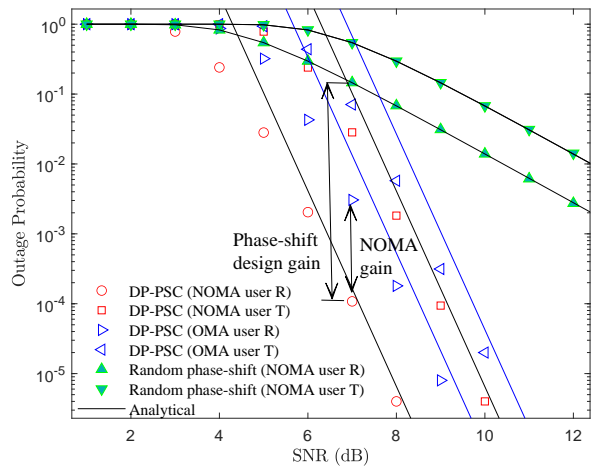}
        \caption{Outage probabilities for NOMA and OMA users for the DP-PSC strategy.}
        \label{compare_out_noma}
    \end{center}
\end{figure}

In Fig.~\ref{compare_out_noma}, we compare the outage probabilities of the users for NOMA and OMA. The STAR-RIS is assumed to employ the DP-PSC strategy. The power allocation factors for the two users are set to $c^2_R = 0.4$ and $c^2_T = 0.6$. As can be observed, for both users, NOMA yields a better outage performance than OMA. Moreover, comparing the DP-PSC strategy with the random PSC strategy, it can be observed that the performance gap between the two PSC strategies (illustrated with the vertical double arrow) increases with the transmit SNR. In contrast, the performance gain of NOMA over OMA in terms of outage probability stays constant in the high SNR regime, which is consistent with \eqref{out_bound1}.

\begin{figure}[h!]
    \begin{center}
        \includegraphics[scale=0.6]{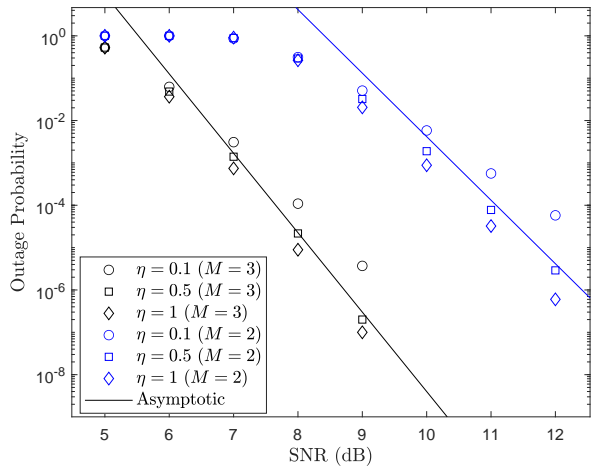}
        \caption{Outage probabilities for diversity preserving strategy with different strengths of direct links.}
        \label{compare_out_d}
    \end{center}
\end{figure}

\subsection{Effect of Direct Links with Different Strengths}
In \eqref{d_pre}, we have shown that with the presence of the direct BS-user links, the diversity orders of both the users for the DP-PSC strategy are $M+1$. In this simulation, we investigate the effects of the direct link strength on the outage probabilities and diversity orders of the users.
Fig.~\ref{compare_out_d} shows the outage probability for the DP-PSC strategy with different direct link strengths. In the legend, $\eta = \mathbb{E}[ |h_d|/|h^\chi_m|]$ denote the power ratio between the expected strength of the direct link and the links through STAR-RIS elements. When $\eta = 0.1$, the direct link is almost negligible. As shown in the figure, the corresponding simulated outage probabilities (plotted with circles) exceed the asymptotic line, meaning that full diversity order is not achieved within the chosen SNR range. However, for $\eta = 0.5$ and $\eta = 1$, it can be observed that full diversity can be achieved despite the change in the strength of the direct link. This observation is consistent with the analytical result derived in \eqref{d_pre}.

\begin{figure}[t!]
    \begin{center}
        \includegraphics[scale=0.6]{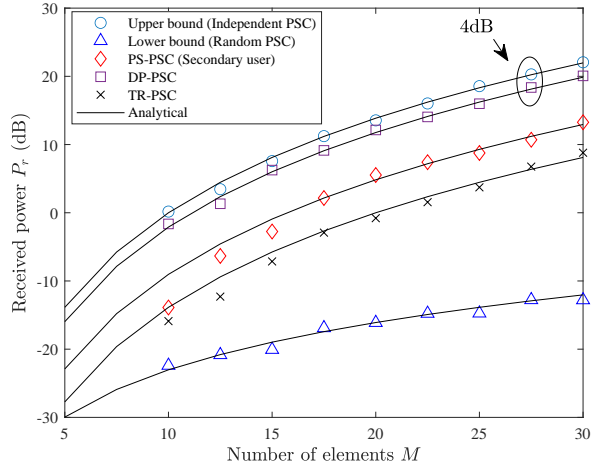}
        \caption{Power scaling laws of users for different PSC strategies.}
        \label{compare_p}
    \end{center}
\end{figure}

\subsection{Power Scaling Laws}
In Fig.~\ref{compare_p}, we investigate the power scaling laws of the proposed PSC strategies. The figure shows the received power in dB (for a reference value of $50$ dBm) versus the number of STAR-RIS elements. For comparison, we also show results for STAR-RIS with the independent phase-shift model and the random PSC strategy, which provide performance upper and lower bounds, respectively. 
As can be observed,
by doubling $M$ from $15$ to $30$, the received powers for the DP-PSC and PS-PSC strategies increase by about $10$ dB, while the power for the random PSC strategy increases only by about $5$ dB.
This is due to their different power scaling laws ($M^2$ versus
$M$), see \eqref{remark_pr}, \eqref{scaling_d_pre}, and \eqref{eq_scaling_random}.
Moreover, the received power for the DP-PSC strategy is only $4$ dB below the upper bound for large numbers of elements. This result can be deduced by comparing \eqref{scaling1} and \eqref{scaling_d_pre} since $10\log_{10}(4/\pi^2)\approx -3.9$ dB.

\section{Conclusions}
In this paper, a correlated T\&R phase-shift model for STAR-RISs was proposed.
Furthermore, considering a STAR-RIS-aided two-user communication system employing OMA and NOMA, three practical PSC strategies were introduced. To evaluate and compare the performance achieved with different STAR-RIS PSC strategies, the asymptotic behavior of the outage probabilities for both OMA and NOMA were derived. 
Moreover, the diversity orders and the power scaling laws for the considered PSC strategies were investigated. 
Our simulations of the beam patterns for different PSC strategies revealed an interesting trade-off between the power of the desired signal and the power of undesired interference. 
Numerical simulations demonstrated that the proposed DP-PSC strategy can achieve the same diversity order and a similar power scaling law as the performance upper bound, with only $4$ dB power degradation.
Numerical results also confirmed the performance gain achieved by NOMA over OMA.

With the proposed correlated phase-shift model, the feasible set of the STAR-RIS phase-shift coefficients becomes non-convex. Thus,
the joint optimization of both the active beamforming at the BS and the correlated T\&R phase shifts of the STAR-RIS constitutes an interesting and challenging topic for future research.

\input{appendix}

\bibliographystyle{IEEEtran}
\bibliography{mybib}

\end{document}

%% file: appendix.tex
\begin{appendices}

\renewcommand{\theequation}{A.\arabic{equation}}
\setcounter{equation}{0}
\section{Proof of \textbf{Theorem~\ref{t_pri}}}\label{ap_b}
Since the expressions for the outage probabilities for OMA and NOMA in \eqref{out_oma}, \eqref{out,r}, and \eqref{out,t} have similar forms, the respective proofs required for \textbf{Theorem~\ref{t_pri}} can be given at the same time.
According to the system model presented in Section~\ref{net_model}, both $|g_m|$ and $|h^R_m|$ follow Rician distributions. Hence, $h_m = \beta^R|g_m||h^R_m|$ is the product of two independent Rician random variables. The PDF of $|h_m|$ can be expressed as follows~\cite{simon2002probability}:
\begin{align}\label{ap_pro_rician}
\begin{split}
    &f_{|h_m|}(x) = \frac{x}{(\beta^R)^2\beta_h\beta_g} e^{-(K_h+K_g)}\\
    &\cdot\sum_{i=0}^\infty\sum_{l=0}^\infty\frac{\big(\frac{\alpha_h\sqrt{x}}{2\beta_h} \big)^{2i}\big(\frac{\alpha_g\sqrt{x}}{2\beta_g} \big)^{2l}}{i!l!\Gamma(i+1)\Gamma(l+1)}
     \big(\frac{\beta_h}{\beta_g}\big)^{\frac{i-l}{2}}K_{i-l}\big(\frac{x}{\beta^R\sqrt{\beta_h\beta_g}}\big),
    \end{split}
\end{align}
where $\alpha^2_{h/g}=\frac{K_{h/g}\Omega_{h/g}}{K_{h/g}+1}$, $\beta_{h/g}=\frac{\Omega_{h/g}}{2(K_{h/g}+1)}$
, $\Gamma(x)$ denotes the Gamma function, and $K_n$ is the modified Bessel function of the second kind.
For the asymptotic behavior, we consider the PDF of $|h_m|$ near the origin using a Taylor series expansion:
\begin{equation}\label{ap_pro_asymp}
 f_{|h_m|}(x) = \frac{e^{-(K^R_h+K_g)}}{(\beta^R)^2\beta_h\beta_g} \cdot x + o(x),
\end{equation}
where $o(\cdot)$ is the little-o notation and $o(f(x))$ denotes a function which is asymptotically smaller than $f(x)$.
The Laplace transform of the asymptotic PDF in \eqref{ap_pro_asymp} is given by:
\begin{equation}
     \mathcal{M}_{|h_m|}(t) \!=\! \mathcal{L}\{f_{|h_m|}\} \!=\! \frac{1}{(\beta^R)^2\beta_h\beta_g} e^{-(K^R_h+K_g)} t^{-2}+o(t^{-2}).
\end{equation}
According to \eqref{hr}, the overall channel of user R is the summation of $M+1$ terms. Using the convolution theorem of the Laplace transform, we obtain the Laplace transform of the asymptotic PDF of $H^R$ as follows: 
\begin{align}\label{l_H_o}
\begin{split}
    \mathcal{M}_{|H^R|}(t) = \mathcal{L}\{f_{|H^R|}\} = \Big(\frac{e^{-(K_h+K_g)}}{(\beta^R)^2\beta_h\beta_g}\Big)^M\cdot \frac{2(K^R_d+1)}{\Omega^R_d}\\ 
    \cdot e^{-K^R_d} 
    t^{-2M-2} +o(t^{-2M-2}).
    \end{split}
\end{align}
Finally, $f_{|H^R|}(x)$ can be obtained by performing the inverse transform term by term. By further substituting $\beta_{h/g}$ with $\frac{\Omega_{h/g}}{K_{h/g}+1}$, we arrive at
 \begin{align}\label{pdf_hr}
 \begin{split}
          f_{|H^R|}(x) &= \frac{2^{M+1}(K^R_h+1)^M(K_g+1)^M(K_d+1)}{(\beta^R)^{2M}\Omega_h^{M}\Omega^M_g\Omega_d} \\ &
          \cdot e^{-MK^R_h-MK_g-K_d}\cdot 
          \frac{x^{2M+1}}{(2M+1)!}.
           \end{split}
      \end{align}
Finally, \textbf{Theorem~\ref{t_pri}} can be proved by integrating the asymptotic PDF in \eqref{pdf_hr} according to \eqref{out,r} and \eqref{out_oma} for NOMA and OMA.
\vspace{-0.1in}
\renewcommand{\theequation}{B.\arabic{equation}}
\setcounter{equation}{0}
\section{Proof of \textbf{Lemma~\ref{lemma_1}}}\label{ap_c}
As shown in Section~\ref{sec_design}, for the secondary user for the PS-PSC strategy and both users for the DP-PSC strategy, the STAR-RIS-aided channel ($h^\chi_s$) can be expressed as follows:
\begin{equation}\label{h_delta_ap}
    h^\chi_s = |g|\beta^\chi \sum_{m=1}^M |h^\chi_m| \exp\{ j(\angle{h^\chi_d}+\delta^\chi_m) \},
\end{equation}
where $\delta^\chi_m$ models the phase error of the $m$th element for user $\chi$. For both considered PSC strategies, this phase error is bounded, i.e., $|\delta^\chi_p-\delta^\chi_q|\leq \Delta^\chi, \forall p,q$. For the PS-PSC strategy, $\Delta^T = \pi$, and for the DP-PSC strategy, $\Delta^R = \Delta^T=\pi/2$.
Hence, as was shown in [17, Corollary 2], 
the PDF of the magnitude of $h^\chi_s$ can be approximated by a Rician distribution:
      \begin{equation}\label{xu_rician_ap}
          f_{|h^\chi_s|}(x)=\frac{x}{\beta^2}e^{-\frac{x^2+\alpha^2}{2\beta^2}}I_0\big(\frac{x\alpha}{\beta}\big),
      \end{equation}
      where
      \begin{align}\label{shape_factors_ap}
          \alpha = M\mathbb{E}[|H_m|]\mathrm{sinc}(\frac{\Delta^\chi}{2}), \ 
          \beta^2 = \frac{M}{2}\mathbb{E}[|H_m|^2][1-\mathrm{sinc}(\Delta^\chi)].
      \end{align}
Thus, for the PS-PSC strategy, the PDF of $|h^\chi_s|$ is obtained by plugging $\Delta^T=\pi$ into \eqref{xu_rician_ap} and (B.3).
\end{appendices}

%% file: final_manuscript.bbl
\begin{thebibliography}{10}
\providecommand{\url}[1]{#1}
\csname url@samestyle\endcsname
\providecommand{\newblock}{\relax}
\providecommand{\bibinfo}[2]{#2}
\providecommand{\BIBentrySTDinterwordspacing}{\spaceskip=0pt\relax}
\providecommand{\BIBentryALTinterwordstretchfactor}{4}
\providecommand{\BIBentryALTinterwordspacing}{\spaceskip=\fontdimen2\font plus
\BIBentryALTinterwordstretchfactor\fontdimen3\font minus
  \fontdimen4\font\relax}
\providecommand{\BIBforeignlanguage}[2]{{%
\expandafter\ifx\csname l@#1\endcsname\relax
\typeout{** WARNING: IEEEtran.bst: No hyphenation pattern has been}%
\typeout{** loaded for the language `#1'. Using the pattern for}%
\typeout{** the default language instead.}%
\else
\language=\csname l@#1\endcsname
\fi
#2}}
\providecommand{\BIBdecl}{\relax}
\BIBdecl

\bibitem{ahead}
M.~{Di Renzo}, A.~{Zappone}, M.~{Debbah}, M.~S. {Alouini}, C.~{Yuen}, J.~{de
  Rosny}, and S.~{Tretyakov}, ``Smart radio environments empowered by
  reconfigurable intelligent surfaces: How it works, state of research, and the
  road ahead,'' \emph{{IEEE} J. Sel. Areas Commun.}, vol.~38, no.~11, pp.
  2450--2525, Nov. 2020.

\bibitem{renzo_diff}
M.~Di~Renzo, K.~Ntontin, J.~Song, F.~H. Danufane, X.~Qian, F.~Lazarakis,
  J.~De~Rosny, D.-T. Phan-Huy, O.~Simeone, R.~Zhang, M.~Debbah, G.~Lerosey,
  M.~Fink, S.~Tretyakov, and S.~Shamai, ``Reconfigurable intelligent surfaces
  vs. relaying: Differences, similarities, and performance comparison,''
  \emph{IEEE Open J. Commun. Soc.}, vol.~1, pp. 798--807, 2020.

\bibitem{liu2020reconfigurable}
Y.~Liu, X.~Liu, X.~Mu, T.~Hou, J.~Xu, M.~Di~Renzo, and N.~Al-Dhahir,
  ``Reconfigurable intelligent surfaces: Principles and opportunities,''
  \emph{IEEE Commun. Surv. Tutor.}, vol.~23, no.~3, pp. 1546--1577, 2021.

\bibitem{xu_star}
J.~Xu, Y.~Liu, X.~Mu, and O.~A. Dobre, ``{STAR-RISs}: Simultaneous transmitting
  and reflecting reconfigurable intelligent surfaces,'' \emph{IEEE Commun.
  Lett.}, vol.~25, no.~9, pp. 3134--3138, May, 2021.

\bibitem{liu_star}
Y.~Liu, X.~Mu, J.~Xu, R.~Schober, Y.~Hao, H.~V. Poor, and L.~Hanzo, ``{STAR}:
  Simultaneous transmission and reflection for $360^\circ$ coverage by
  intelligent surfaces,'' \emph{accepted for publication in {IEEE} Wireless
  Commun., Available: arXiv:2103.09104}, 2021.

\bibitem{doc}
\BIBentryALTinterwordspacing
N.~DOCOMO. ``{DOCOMO} conducts world’s first successful trial of transparent
  dynamic metasurface''. [Online]. Available:
  \url{www.nttdocomo.co.jp/english/info/media\_center/pr/2020/0117\_00.html}
\BIBentrySTDinterwordspacing

\bibitem{zhang2021intelligent}
H.~Zhang, S.~Zeng, B.~Di, Y.~Tan, M.~Di~Renzo, M.~Debbah, L.~Song, Z.~Han, and
  H.~Poor, ``Intelligent omni-surfaces for full-dimensional wireless
  communications: Principle, technology, and implementation,'' \emph{arXiv
  preprint arXiv:2104.12313}, 2021.

\bibitem{xu_vtmag}
J.~Xu, Y.~Liu, X.~Mu, J.~T. Zhou, L.~Song, H.~V. Poor, and L.~Hanzo,
  ``Simultaneously transmitting and reflecting ({STAR}) intelligent
  omni-surfaces, their modeling and implementation,'' \emph{arXiv preprint
  arXiv:2108.06233}, 2021.

\bibitem{mu2021simultaneously}
X.~Mu, Y.~Liu, L.~Guo, J.~Lin, and R.~Schober, ``Simultaneously transmitting
  and reflecting ({STAR}) {RIS} aided wireless communications,'' \emph{{IEEE}
  Trans. Wireless Commun.}, Early Access, 2021, doi: 10.1109/TWC.2021.3118225.

\bibitem{star_coverage}
C.~Wu, Y.~Liu, X.~Mu, X.~Gu, and O.~A. Dobre, ``Coverage characterization of
  {STAR-RIS} networks: {NOMA} and {OMA},'' \emph{IEEE Commun. Lett.}, vol.~25,
  no.~9, pp. 3036--3040, 2021.

\bibitem{basar}
E.~Basar, M.~Di~Renzo, J.~De~Rosny, M.~Debbah, M.-S. Alouini, and R.~Zhang,
  ``Wireless communications through reconfigurable intelligent surfaces,''
  \emph{IEEE Access}, vol.~7, pp. 116\,753--116\,773, 2019.

\bibitem{huang}
C.~Huang, A.~Zappone, G.~C. Alexandropoulos, M.~Debbah, and C.~Yuen,
  ``Reconfigurable intelligent surfaces for energy efficiency in wireless
  communication,'' \emph{IEEE Trans. Wireless Commun.}, vol.~18, no.~8, pp.
  4157--4170, 2019.

\bibitem{mu2019exploiting}
X.~{Mu}, Y.~{Liu}, L.~{Guo}, J.~{Lin}, and N.~{Al-Dhahir}, ``Exploiting
  intelligent reflecting surfaces in {NOMA} networks: Joint beamforming
  optimization,'' \emph{{IEEE} Trans. Wireless Commun.}, vol.~19, no.~10, pp.
  6884--6898, 2020.

\bibitem{sensing}
J.~Hu, H.~Zhang, B.~Di, L.~Li, K.~Bian, L.~Song, Y.~Li, Z.~Han, and H.~V. Poor,
  ``Reconfigurable intelligent surface based rf sensing: Design, optimization,
  and implementation,'' \emph{{IEEE} J. Sel. Areas Commun.}, vol.~38, no.~11,
  pp. 2700--2716, 2020.

\bibitem{rui_couple}
S.~Abeywickrama, R.~Zhang, Q.~Wu, and C.~Yuen, ``Intelligent reflecting
  surface: Practical phase shift model and beamforming optimization,''
  \emph{IEEE Trans. Commun.}, vol.~68, no.~9, pp. 5849--5863, 2020.

\bibitem{9319694}
G.~Gradoni and M.~Di~Renzo, ``End-to-end mutual coupling aware communication
  model for reconfigurable intelligent surfaces: An electromagnetic-compliant
  approach based on mutual impedances,'' \emph{IEEE Wireless Commun. Lett.},
  vol.~10, no.~5, pp. 938--942, 2021.

\bibitem{xu2020novel}
J.~Xu and Y.~Liu, ``A novel physics-based channel model for reconfigurable
  intelligent surface-assisted multi-user communication systems,'' \emph{IEEE
  Trans. Wireless Commun.}, Early Access, 2021, doi:10.1109/TWC.2021.3102887.

\bibitem{vlc}
A.~R. Ndjiongue, T.~M.~N. Ngatched, O.~A. Dobre, and H.~Haas, ``Toward the use
  of re-configurable intelligent surfaces in {VLC} systems: Beam steering,''
  \emph{{IEEE} Wireless Commun.}, vol.~28, no.~3, pp. 156--162, 2021.

\bibitem{danufane2020path}
F.~H. Danufane, M.~Di~Renzo, J.~de~Rosny, and S.~Tretyakov, ``On the path-loss
  of reconfigurable intelligent surfaces: An approach based on green's theorem
  applied to vector fields,'' \emph{arXiv preprint arXiv:2007.13158}, 2020.

\bibitem{liu2021simultaneously}
Y.~Liu, X.~Mu, R.~Schober, and H.~V. Poor, ``Simultaneously transmitting and
  reflecting {(STAR)-RISs}: A coupled phase-shift model,'' \emph{arXiv preprint
  arXiv:2110.02374}, 2021.

\bibitem{old}
I.~V{\'a}g{\'o}, ``On the interface and boundary conditions of electromagnetic
  fields,'' \emph{Period. Polytech. Elec. Eng.}, vol.~38, no.~2, pp. 79--94,
  1994.

\bibitem{rothwell2018electromagnetics}
E.~J. Rothwell and M.~J. Cloud, \emph{Electromagnetics}.\hskip 1em plus 0.5em
  minus 0.4em\relax CRC Press, 2018.

\bibitem{zhu2014dynamic}
B.~O. Zhu, K.~Chen, N.~Jia, L.~Sun, J.~Zhao, T.~Jiang, and Y.~Feng, ``Dynamic
  control of electromagnetic wave propagation with the equivalent principle
  inspired tunable metasurface,'' \emph{Sci. Rep.}, vol.~4, no.~1, pp. 1--7,
  2014.

\bibitem{book}
T.~L. Marzetta and H.~Q. Ngo, \emph{Fundamentals of massive {MIMO}}.\hskip 1em
  plus 0.5em minus 0.4em\relax Cambridge University Press, 2016.

\bibitem{liu_enhancing}
Y.~{Liu}, Z.~{Qin}, M.~{Elkashlan}, Y.~{Gao}, and L.~{Hanzo}, ``Enhancing the
  physical layer security of non-orthogonal multiple access in large-scale
  networks,'' \emph{{IEEE} Trans. Wireless Commun.}, vol.~16, no.~3, pp.
  1656--1672, 2017.

\bibitem{onebit}
T.~Wang, G.~Chen, J.~P. Coon, and M.-A. Badiu, ``Study of intelligent
  reflective surface assisted communications with one-bit phase adjustments,''
  in \emph{Proc. {IEEE} Global Commun. Conf. ({GLOBECOM})}, 2020, pp. 1--6.

\bibitem{simon2002probability}
M.~K. Simon, \emph{Probability distributions involving Gaussian random
  variables: A handbook for engineers and scientists}.\hskip 1em plus 0.5em
  minus 0.4em\relax Springer, 2002.

\end{thebibliography}
